\documentclass[a4paper,12pt,preprint,amsmath,showpacs,nofootinbib,groupedaddress,tightenlines]{revtex4-1}
\usepackage{color}
\usepackage{slashed}
\usepackage{graphicx}   
\usepackage{slashed}
\usepackage{color}
\usepackage[lofdepth,lotdepth]{subfig}
\usepackage{psfrag}
\usepackage{eufrak}
\usepackage{accents}

\newcommand{\dd}{\mathrm{d}}

\newcommand{\als}{\alpha_s}
\newcommand{\ep}{\epsilon}

\newcommand{\nn}{\nonumber}

\newcommand{\ws}{\widetilde{S}}

\begin{document}

\preprint{MITP/14-055, SLAC-PUB-15982}

\title{The two-loop soft function for heavy quark\\ pair production at future linear colliders}

\author{Andreas von Manteuffel}
\author{Robert M. Schabinger}
\affiliation{The PRISMA Cluster of Excellence and
Mainz Institute of Theoretical Physics,
Johannes Gutenberg Universit\"at,
55099 Mainz, Deutschland}
\author{Hua Xing Zhu}
\affiliation{SLAC National Accelerator Laboratory, Stanford University, Stanford, CA 94309, USA\\}

\begin{abstract}
\noindent
We report on the calculation of the threshold soft function for heavy quark pair production in $e^+ e^-$ annihilation at two-loop order.
Our main result is a generalization of the familiar Drell-Yan threshold soft function to the case of non-zero primary quark mass.
We set up a framework based on the method of differential equations which allows for the straightforward calculation of the bare soft function to arbitrarily high orders in the dimensional regularization parameter.
Remarkably, we find that we can obtain the bare two-loop Drell-Yan soft function from the heavy quark soft function to the order in epsilon required for a two-loop calculation by making simple replacements.
We expect that our results will be of use, both as an important input for precision physics calculations at linear colliders and, more formally,
as a first step towards a better understanding of the connection between vacuum matrix elements of massive soft Wilson lines and vacuum matrix elements of massless soft Wilson lines.
\end{abstract}

\maketitle
\allowdisplaybreaks
\section{I\lowercase{ntroduction}}
\label{sec:intro}

Soft and collinear factorization formulas for infrared-safe observables in Quantum Chromodynamics (QCD) often
involve soft functions. Up to normalization, a soft function for a particular QCD observable is nothing but the expression for that observable in the soft limit.
The calculation of the soft function for a given observable is typically much simpler than the analogous full QCD calculation but can nevertheless provide useful physics
insights which might otherwise be cumbersome to extract analytically~\cite{hep-ph/9808389,hep-ph/0102227,hep-ph/0102241}.

Recently, due to an ever-increasing experimental demand for precise theoretical predictions, numerous higher-order soft function calculations have been carried out. 
Most of these higher-order computations are at $\mathcal{O}\left(\als^2\right)$ in perturbative QCD and concern observables with soft limits which are completely described by massless (light-like) soft Wilson lines. 
Notable examples include various soft functions for jet mass observables in $e^+e^-$ annihilation~\cite{1105.3676,1105.4560,1105.4628,1112.3343,1309.3560},
the soft function for $B \to X_s \gamma$ decay~\cite{hep-ph/0501257,hep-ph/0512208},
the transverse momentum-dependent soft function for Drell-Yan production~\cite{1105.5171}, the soft function for electroweak gauge boson production at large transverse momentum~\cite{1201.5572}, and
the soft function for the invariant mass distribution of highly boosted $t \bar{t}$ pairs~\cite{1207.4798}. Very recently, remarkable progress has
been made towards the calculation of soft corrections at $\mathcal{O}\left(\als^3\right)$~\cite{1302.4379,1309.4391,1309.4393,1311.1425,1312.1296,1404.5839}. As
a result, threshold predictions for both Higgs and Drell-Yan production at hadron colliders are now available at next-to-next-to-next-to-leading order~\cite{1403.4616,1404.0366}.

On the other hand, much less attention has been given to higher-order soft function calculations which involve squared matrix elements of massive (time-like) Wilson lines.
In fact, only the soft function for secondary production of a massive quark-antiquark pair in $e^+ e^-$ annihilation~\cite{1309.6251}
and the soft function for threshold production of a massive color octet at hadron colliders~\cite{1311.2541} have been calculated to second order in the strong coupling constant.
In particular, only one-loop, next-to-leading order~(NLO) QCD corrections are currently available for the threshold top quark pair
production soft function at hadron colliders~\cite{1003.5827}. It is natural to ask whether the calculation of the complete two-loop, next-to-next-to-leading order~(NNLO) corrections to the threshold soft
function for top quark pair production at hadron colliders can be carried out because it is
an important ingredient for both next-to-next-to-next-to-leading threshold logarithm resummation and approximate fully differential NNLO calculations~\cite{1406.0386,1407.2532}. 
In this work, we take a first step towards our ultimate goal of understanding the parton-initiated processes relevant for hadron collider physics by calculating the threshold soft function for heavy quark pair production in $e^+ e^-$ annihilation.

The motivation for this calculation is manifold. 
First of all, the master integrals that we encounter during the course of our $e^+ e^-$ calculation will surely also show up when we calculate the two-loop $p p \to t \bar{t}$ soft function.
Furthermore, the calculation performed in this paper can and will~\cite{1408.5150} be used to construct a Monte Carlo program for the computation of fully differential NNLO
QCD corrections to the continuum production of heavy quark pairs at $e^+ e^-$ colliders. So far, no similar program has appeared which furnishes fully differential NNLO predictions for continuum $b \bar{b}$ and
$t \bar{t}$ production in $e^+ e^-$ annihilation. 
Finally, the structure of the result obtained is interesting in its own right because, to date, no complete two-loop, massive soft function calculation of comparable complexity has been carried out.

As alluded to above, the most immediate application of our results will be to new studies of heavy quark pair production in $e^+ e^-$ annihilation at energies far above the kinematic threshold.
A good understanding of such processes, both theoretically and experimentally, is of considerable interest.
For instance, the bottom-quark forward-backward asymmetry at the $Z$ peak has long been known to provide one of the most precise methods for the extraction of the electroweak mixing angle, $\theta_W$~\cite{hep-ex/0509008}. 
Furthermore, at the proposed International Linear Collider, where a sustained run at a center-of-mass energy of 1 TeV has been discussed, 
precise fully differential continuum top-quark pair production calculations will be essential
in order to obtain reliable theoretical predictions for important high energy collider observables like the top-quark forward-backward asymmetry.

Due to the presence of the heavy quark mass, higher-order perturbative QCD corrections to the process
\begin{eqnarray}
  e^+ e^- \to \gamma^*/ Z^* \to Q\bar{Q} + X
\end{eqnarray}
are difficult to calculate, even at NNLO. This is reflected in the fact that, to date, all physics predictions at NNLO have been made in some sort of approximation scheme.
A number of studies of the total cross section have been carried out over the years, both at the kinematic threshold~\cite{hep-ph/9712222,hep-ph/9712302,hep-ph/9801397,hep-ph/0001286}
and at extremely high energies~\cite{Gorishnii:1986pz,PHLTA.B248.359,NUPHA.B432.337,hep-ph/9704222}. 
For more differential observables such as the forward-backward asymmetry, the situation is similar. 
The heavy quark forward-backward asymmetry has been known to NNLO in the massless approximation for some time~\cite{NUPHA.B391.3,hep-ph/9809411,hep-ph/9905424} and, more recently,
the full mass dependence was worked out for the purely virtual NNLO corrections~\cite{hep-ph/0604031}.
At the present time, no fully differential calculation, exact to second order in QCD perturbation theory, exists in the literature.

The calculation of infrared-safe observables at NNLO is also complicated by the fact that, for a given observable, there are always three independent components which must be calculated and then consistently combined together;
in order to obtain a finite and physically meaningful result, one requires two-loop, purely virtual corrections, one-loop, real-virtual corrections with one additional unresolved parton in the final state,
and real-real corrections with two additional unresolved partons in the final state.
Over the years, a number of important steps have been taken towards the computation of fully differential NNLO QCD corrections to $e^+e^-\to Q\bar{Q}$ observables.
The two-loop, purely virtual corrections were first computed in a series of papers~\cite{hep-ph/0406046,hep-ph/0412259,hep-ph/0504190} and later confirmed by another group~\cite{0905.1137}.
When at least one parton is resolved, the combination of the real-virtual
and real-real corrections are given by the well-known NLO QCD corrections to $e^+e^-\to Q \bar{Q} + {\rm jet}$~\cite{hep-ph/9703358,hep-ph/9705295,hep-ph/9708350,hep-ph/9709360,hep-ph/9905276}. In fact,
the only ingredient missing is a framework for the consistent combination of the various components mentioned above.
Important progress has been made on this front by a number of authors, extending the antenna subtraction method~\cite{hep-ph/0505111} to deal
with massive, colored, final state particles~\cite{1102.2443,1105.0530,1309.6887}. In the meantime, a promising new method~\cite{1005.0274} has been developed which makes use of ideas from Frixione-Kunszt-Signer
subtraction~\cite{NUPHA.B467.399,NUPHA.B486.189} and sector decomposition~\cite{NUPHA.B585.741,PHRVA.D69.076010,NUPHA.B693.134}.

An alternative and perhaps even more straightforward approach to such calculations is to simply extend to NNLO the phase space slicing method used in the original NLO QCD
calculation of the $e^+e^- \to Q \bar Q$ process~\cite{Jersak:1981sp}. In fact, a variant of this approach based on dispersion relations has already been successfully applied to compute
the full light quark flavor-dependent contributions at NNLO~\cite{hep-ph/9505262,hep-ph/9707496}.
To fix some notation and help clarify the role played by the $e^+e^- \to Q \bar Q$ soft function,
we now give a brief explanation of the method in the context of a simple example.\footnote{We refer the reader to reference~\cite{1408.5150} for an unabridged exposition of the NNLO phase space slicing method.}

In the region of phase space where the energy of the QCD radiation off of the heavy quarks is small,
heavy quark effective theory~\cite{PHLTA.B234.511,NUPHA.B339.253,PHLTA.B237.527,PHLTA.B240.447} implies that $Q \bar{Q}$  differential distributions factorize. We therefore have, for instance, the factorization formula
\begin{equation}
\label{eq:factformula}
\frac{\dd \sigma^{Q \bar{Q}}}{\dd \cos\theta} = \frac{\dd \sigma^{Q \bar{Q}}_0}{\dd \cos\theta}\, H^{Q \bar{Q}}\left(x, \mu\right) \int_0^{2 E_{cut}} \dd \lambda \,S^{Q \bar{Q}}\left(x, \lambda, \mu\right)
~~+~~\mathcal{O}\left(E_{cut}/\sqrt{s}\right)\,,
\end{equation}
relevant to the computation of the $Q \bar{Q}$ forward-backward asymmetry (see {\it e.g.} \cite{hep-ph/0604031}). In the above, $\dd \sigma^{Q \bar{Q}}_0/\dd \cos\theta$ is the leading order cross section for
$Q\bar{Q}$ production differential in the cosine of the polar angle, $\theta$, between the beam line and the final state heavy quark, $\lambda$ is (for essentially historical reasons)
twice the energy of the soft radiation off of the $Q\bar{Q}$ pair, and $\mu$ is the factorization scale. The kinematical variable $x$ appearing in the factorization formula is defined in terms of the heavy quark mass,
$m_Q$, and the center-of-mass energy, $s$, as
\begin{equation}
\label{eq:x}
x = \frac{1 - \sqrt{1-\frac{4 m_Q^2}{s}}}{1 + \sqrt{1-\frac{4 m_Q^2}{s}}}\,.
\end{equation}
The hard function, $H^{Q \bar{Q}}\left(x, \mu\right)$, encodes the purely virtual corrections and the soft function, $S^{Q \bar{Q}}\left(x, \lambda, \mu\right)$, encodes the soft radiative corrections at threshold.
Finally, the factorization formula depends on an energy scale, $E_{cut}$,  below which the power corrections in $E_{cut}/\sqrt{s}$ become negligible and, as a result,
the full differential distribution is well-approximated by Eq.\ (\ref{eq:factformula}). Now, continuing with our example, the phase space slicing method would first 
employ factorization formula (\ref{eq:factformula}) to calculate the differential cross section when the total energy of the soft radiation is less than or equal to $E_{cut}$
and then employ a Monte Carlo calculation of  $e^+e^-\to Q \bar{Q} + {\rm jet}$ to calculate the differential cross section
for jet energies greater than or equal to $E_{cut}$.
The idea is to obtain a result which is approximately independent of the cutoff by taking $E_{cut}$ small enough to sufficiently suppress power corrections to Eq.\ (\ref{eq:factformula})
but at the same time large enough to avoid numerical instabilities in the Monte-Carlo code for $e^+e^-\to Q \bar{Q} + {\rm jet}$ for jet energies of order $E_{cut}$.
In a closely related context, stability with respect to variations in the cutoff parameter was demonstrated in reference \cite{1210.2808}.

This article is organized as follows. In Section~\ref{sec:2}, we write down the operator definition of the $e^+ e^-\to Q\bar{Q}$ soft function and discuss its renormalization group~(RG) evolution. 
In Section~\ref{sec:oneloopsol}, we exhibit the one-loop heavy quark soft function to all orders in the dimensional regularization parameter.
The remainder of the third section is devoted to the technical details of our two-loop calculation, 
first in Section \ref{sec:realvirtual} for the real-virtual contributions and then in Section \ref{sec:realreal} for the real-real contributions.
Due to their length, we chose to put our results for the real-virtual and real-real contributions to the bare two-loop soft function in ancillary files included with the arXiv submission of this paper.
It is also worth pointing out that readers interested only in our final result for the renormalized two-loop soft function may skip directly from Section~\ref{sec:2} to Section~\ref{sec:4} without significant loss of continuity.
In Section~\ref{sec:4}, we begin by presenting the final result of our calculation in a concise way. In order to check that we have not made any calculational errors, we then discuss, one-by-one,
all available consistency checks on the result. This naturally leads into a discussion of the light-like limit of the heavy quark soft function and our novel way of understanding it.
With no additional input whatsoever,
we found that we could predict the $\ep$ expansion of the bare two-loop $e^+ e^- \rightarrow q\bar{q}$ or Drell-Yan soft function to $\mathcal{O}(\ep^i)$ by expanding our bare two-loop heavy quark soft function to $\mathcal{O}(\ep^{i+1})$
and then making certain replacements in the resulting expressions.
Besides a precise statement of the connection between these two seemingly rather different bare soft functions,
we discuss another, similar correspondence between bare soft functions defined using a hemisphere jet algorithm and point out that,
in fact, the various relations we have observed cannot follow from known factorization properties of the massive soft functions we consider.
Finally, a brief summary and outlook is given in Section~\ref{sec:5}.

\section{D\lowercase{efinition of the soft function and its} RG \lowercase{evolution}}
\label{sec:2}

The soft function can be written as the square of a time-ordered matrix element of
two semi-infinite Wilson line operators,
\begin{eqnarray}
\label{eq:operatordef}
 S^{Q \bar{Q}}\left(x, \lambda, \mu\right) = \frac{1}{N_c} \sum_{\mbox{\tiny $X_S$ }}
\langle 0 | T\Big\{Y_{v}^\dagger Y_{\bar{v}}\Big\} \delta\Big(\lambda - \hat{\bf P}^0\Big) | X_S\rangle\langle X_S | T\Big\{Y^\dagger_{\bar{v}} Y_{v}\Big\} | 0 \rangle.
\end{eqnarray}
The soft function defined in Eq.\ (4) depends on the kinematical parameter $x$ defined in Eq.\ (3), twice the energy of the soft QCD radiation, $\lambda$, and the factorization scale, $\mu$.
The soft function also depends implicitly on the renormalized strong coupling constant, $\als(\mu)$, and the number of colors, $N_c$. The summation is over all possible soft parton final
states, $|X_S\rangle$. The operator $\hat{\bf P}^0$ acts on the final state $|X_S\rangle$ according to
\begin{equation}
 \hat{\bf P}^0|X_S\rangle = 2 E_{X_S} |X_S\rangle,
\end{equation}
where $E_{X_S}$ is the energy of the soft radiation in final state $X_S$. The Wilson line operators, $Y_{v}^\dagger$ and $Y_{\bar{v}}$, are respectively defined as out-going, path-ordered $\left(\mathbf{P}\right)$
 and anti-path-ordered $\left(\overline{\mathbf{P}}\right)$ exponentials \cite{hep-ph/0412110},
\begin{eqnarray}
  Y^\dagger_{v}(y) &=& \mathbf{P} \exp \left( i g \int^\infty_0 \dd r\, v \cdot A(v\, r+y) \right)
\nn
\\
  Y_{\bar{v}}(y) &=& \overline{\mathbf{P}} \exp \left( -i g \int^\infty_0  \dd r\, \bar{v} \cdot A(\bar{v}\, r+y) \right).
\end{eqnarray}
In the above, $A_\mu = A_{\mu}^a T^a$, where the $T^a$ are fundamental $\mathfrak{su}(N_c)$ matrices, $v$ is the velocity vector of $Q$, and
$\bar{v}$ is the velocity vector of $\bar{Q}$. For a generic four-vector, $k^\mu$, it is a straightforward exercise to show that
\begin{eqnarray}
v \cdot k &=& \frac{1}{1+x} k^+ + \frac{x}{1+x} k^-\\
\bar{v} \cdot k &=& \frac{x}{1+x} k^+ + \frac{1}{1+x} k^-
\end{eqnarray}
given the usual definitions $k^+ = k^0 + k^3$ and $k^- = k^0 - k^3$. From the above it is clear that the $e^+ e^- \to q \bar{q}$ and, by virtue of the time-reversal invariance of QCD, Drell-Yan
soft functions~\cite{hep-ph/9808389} can be computed from Eq.\ (4) in the $x \rightarrow 0$ limit.

In the perturbative regime, the soft function can be calculated order-by-order in $\als(\mu)$. Let us begin by discussing the bare soft function. Already at leading order in $\als$, the soft function
develops ultraviolet~(UV) divergences\footnote{The infrared~(IR) divergences
cancel between the purely virtual contributions to the squared matrix element and those contributions which have some number of soft partons in the final state. Furthermore, the purely virtual contributions to the squared matrix element of 
the Wilson line operators are all scaleless and thus vanish in
dimensional regularization. It therefore follows that this infrared cancellation amounts to a
conversion of IR poles to UV poles.} and, therefore, non-trivial
dependence on $\mu$. By dimensional analysis, we can write the loop expansion of the bare soft function as\footnote{The bare expansion coefficients of the soft function, $S^{(L)}_{Q\bar{Q}}(x,\ep)$, are defined 
in the obvious way by setting the parameters $\lambda$ and $\mu$ in Eq.\ (\ref{eq:operatordef}) to one. This approach is convenient because, to all orders in the loop expansion, the dependence on the scales $\lambda$ and $\mu$
is trivially determined by power counting.}
\begin{align}
  S_{\rm bare}^{Q \bar{Q}} (x, \lambda, \mu) = \delta(\lambda) + \frac{\als S_\ep}{4\pi}
  \frac{\mu^{2\ep}}{\lambda^{1+2\ep}}S^{(1)}_{Q\bar{Q}}(x,\ep) 
+ \left(\frac{\als S_\ep}{4\pi} \right)^2
  \frac{\mu^{4\ep}}{\lambda^{1+4\ep}}S^{(2)}_{Q\bar{Q}}(x,\ep) + \mathcal{O}\left(\als^3\right),
\end{align}
where the $S^{(n)}_{Q\bar{Q}}(x,\ep)$ are coefficient functions depending
on both $x$ and the parameter of dimensional regularization, $\ep=(4-d)/2$. The renormalization of the soft function is most
easily carried out in Laplace space. As usual, the Laplace transform is taken with respect to $\lambda$, 
\begin{eqnarray}
\ws_{\rm bare}^{Q \bar{Q}} (x, L) =   \int^\infty_0 \dd\lambda \exp\left(
  \frac{-\lambda}{e^{\gamma_E}  \kappa} \right) S_{\rm bare}^{Q \bar{Q}} (x, \lambda, \mu),
\end{eqnarray}
where $L= \ln (\kappa/\mu)$ and $\gamma_E=0.5772\dots$ is the Euler constant.

The bare soft function in Laplace space has the simple structure
\begin{eqnarray}
\label{eq:LaplaceExp}
  \ws_{\rm bare}^{Q \bar{Q}} (x, L) &=& 
  1 + \frac{\als S_\ep}{4\pi}\ws^{(1)}_{Q\bar{Q}}(x, L, \ep) + \left(\frac{\als S_\ep}{4\pi}\right)^2\ws^{(2)}_{Q\bar{Q}}(x, L, \ep) + \mathcal{O}\left(\als^3\right)
 \nn\\
  &=& 1 + \frac{\als S_\ep}{4\pi}e^{-2\ep \gamma_E}\Gamma(-2\ep)e^{-2\ep L}S^{(1)}_{Q\bar{Q}}(x,\ep)
\nonumber\\  
  &&+  \left(\frac{\als S_\ep}{4\pi}\right)^2 e^{-4\ep \gamma_E}\Gamma(-4\ep) e^{-4\ep L}S^{(2)}_{Q\bar{Q}}(x,\ep) + \mathcal{O}\left(\als^3\right)
\end{eqnarray}
and it requires both charge and operator renormalization. Charge renormalization is carried out in the 
$\overline{\rm MS}$ scheme by replacing the bare strong coupling constant with its renormalized
counterpart, 
\begin{eqnarray}
 \alpha_s S_\ep  = \alpha_s(\mu) \left[  1 - \frac{\alpha_s(\mu)}{4\pi} \frac{\beta_0}{\ep} + \mathcal{O}\left(\alpha^2_s(\mu)\right) \right],
\end{eqnarray}
where $S_\ep = \left(4 \pi e^{-\gamma_E}\right)^\ep$, and 
\begin{eqnarray}
  \beta_0  = \frac{11}{3} C_A - \frac{4}{3} n_f T_F
\end{eqnarray}
is the leading order QCD beta function with $n_f$ light flavors. In this paper, $T_F=1/2$, $C_A=3$, and $n_f=5$.

An additional operator renormalization is typically necessary whenever one studies composite field operators. In this case, the appearance of additional UV divergences is
related to the fact that the product of Wilson line operators in the definition of the soft function has a
cusp singularity at the coordinate origin. This cusp
singularity can be removed by a simple multiplicative renormalization,
\begin{eqnarray}
  \ws^{Q \bar{Q}} (x, L) = Z_S(x) \,\ws_{\rm bare}^{Q \bar{Q}} (x, L).
\end{eqnarray}
After charge renormalization has been carried out, the operator renormalization constant of the soft function is fixed at each order in the renormalized coupling
by insisting that it absorb any remaining UV divergences. It admits an
expansion in $\alpha_s(\mu)$ of the form
\begin{eqnarray}
  Z_S(x) = 1 +\frac{\alpha_s(\mu)}{4\pi} Z_S^{(1)}(x) + \left(
    \frac{\alpha_s(\mu)}{4\pi}\right)^2 \left(\frac{\left(Z_S^{(1)}(x)\right)^2}{2} +
    Z_S^{(2)}(x)  \right) + \mathcal{O}(\alpha_s(\mu)).
\label{eq:opz}
\end{eqnarray}
Later in this section, we will give expressions for $Z_S^{(1)}(x)$ and $Z_S^{(2)}(x)$ in terms of known universal quantities.
At this stage, it is worth emphasizing that, unless otherwise stated, we keep the dependence on $x$ exact in all of our calculations. In particular, we make frequent use of the relations
\begin{equation}
v^2 = \bar{v}^2 = \frac{4 x}{(1+x)^2} \qquad \qquad v\cdot \bar{v} = \frac{2(1+x^2)}{(1+x)^2}.
\end{equation}

The renormalized soft function in Laplace space obeys a well-known evolution
equation~\cite{NUPHA.B283.342} of the form
\begin{equation}
\label{eq:evolution}
  \frac{d\ws^{Q \bar{Q}}(x, L)}{d\ln(\mu)} = -\gamma^s\left(x\right) \ws^{Q \bar{Q}}(x, L).
\end{equation}
In Eq.\ (\ref{eq:evolution}) above, $\gamma^s(x)$ is the cusp anomalous dimension for massive quarks. As usual, the cusp anomalous dimension admits an expansion in the renormalized coupling,
\begin{eqnarray}
  \gamma^s (x) = \frac{\als(\mu)}{4\pi} \gamma^s_0(x) + \left(
    \frac{\als(\mu)}{4\pi} \right)^2\gamma^s_1(x) + \mathcal{O}\left(\als(\mu)^3\right).
\end{eqnarray}
It turns out that both $\gamma^s_0(x)$ and $\gamma^s_1(x)$ can be written in a natural and compact way using harmonic polylogarithms~\cite{hep-ph/9905237} of argument $x$. 

Let us state explicitly that our harmonic polylogarithms, $G(w_1,\dots,w_n;x)$, differ slightly from those of reference~\cite{hep-ph/9905237}. Following~\cite{1309.3560}, we begin by setting $G(;x) = 1$ for all $x$. 
Harmonic polylogarithms with $n$ weights are then defined as follows. Consider a set of $n$ integers, $\{w_1,\ldots,w_n\}$, drawn from the set $\{-1,0,1\}$. 
If at least one of $\{w_1,\ldots,w_n\}$ is different from zero, the weight $n$ harmonic polylogarithm with weight vector $\{w_1,\ldots,w_n\}$ is defined recursively as
\begin{equation}
G(w_1,\dots,w_n;x) = \int_0^x {dt \over t - w_1} G(w_2,\dots,w_n;t).
\end{equation}
If, on the other hand, $w_i = 0$ for all $i$, then the weight $n$ harmonic polylogarithm with weight vector $\{w_1,\ldots,w_n\}$ is given by
\begin{equation}
G(0,\dots,0;x) = \frac{1}{n!}\ln^n(x)\,.
\end{equation}
In this work, we found it useful at various stages to perform high precision numerical evaluations of our harmonic polylogarithms using the implementation~\cite{hep-ph/0410259} available in the {\tt GiNaC} computer algebra system~\cite{cs.sc/0004015}.

We are now in a position to write down both $\gamma^s_0(x)$ and $\gamma^s_1(x)$ explicitly~\cite{NUPHA.B283.342,hep-ph/0406046,0903.2561,0904.1021}. In our notation, we have
\begin{eqnarray}
&&\gamma^s_0(x) = -8 C_F \left[1+\frac{1+x^2}{1-x^2}G(0;x)\right]
\label{eq:anomalous1}
\\
&&\gamma^s_1(x) = \frac{160}{9} C_F n_f T_F \left[1+\frac{1+x^2}{1-x^2}G(0;x)\right]
\label{eq:anomalous2}
\\ &&
+ C_A C_F \left[
-\frac{392}{9}+\frac{8 \pi^2(1 + 9 x^2)}{3(1-x^2)}+16\frac{\left(1+x^2\right)^2}{\left(1-x^2\right)^2} \Big(2 G(0;x) \Big(G(0,-1;x)+G(0,1;x)\Big)\nn
\right.\\&&\left.
-4 G(0,0,-1;x)-4 G(0,0,1;x)-\zeta (3)\Big)+32\frac{1+x^2}{1-x^2} \bigg(G (0;x) \left( G(-1;x)+G(1;x)-\frac{67}{36}\right)\nn
\right.\\&&\left.
-G(0,-1;x)-G(0,1;x)\bigg)-\frac{32 x^2(1+x^2)}{3\left(1-x^2\right)^2}G^3(0;x)-\frac{32 x^2}{1-x^2} G^2(0;x)\nn
\right.\\&&\left.
+\frac{8 \pi ^2 \left(1+x^2\right)\left(1+9 x^2\right)}{3\left(1-x^2\right)^2}G(0;x)
\right].\nn
\end{eqnarray}
Eqs.\ (\ref{eq:anomalous1}) and (\ref{eq:anomalous2}) were derived by first extracting the cusp anomalous dimensions in the space-like region from reference~\cite{hep-ph/0406046}
and then analytically continuing the obtained expressions to the time-like region. 

Evolution equation (\ref{eq:evolution}) determines the form of the renormalization constants $Z_S^{(1)}(x)$ and $Z_S^{(2)}(x)$
in terms of universal quantities defined above:
\begin{eqnarray}
\label{eq:Z1}
Z_S^{(1)}(x) &=& \frac{\gamma_0^s(x)}{2\ep}\\
\label{eq:Z2}
Z_S^{(2)}(x) &=& -\frac{\beta_0 \gamma_0^s(x)}{4 \ep^2} + \frac{\gamma_1^s(x)}{4 \ep}.
\end{eqnarray}
Eqs.\ (\ref{eq:Z1}) and (\ref{eq:Z2}) are derived by plugging Eq.\ (\ref{eq:LaplaceExp}) into Eq.\ (\ref{eq:evolution}) and taking care to keep the $\ep$ dependence exact at all stages.
It is also possible to solve Eq.\ (\ref{eq:evolution}) in the $\ep \rightarrow 0$ limit
up to $\mu$-independent constants and terms of higher order in $\als(\mu)$,
\begin{eqnarray}
  \ws^{Q \bar{Q}}(x, L) &=& 1 + \frac{\als(\mu)}{4\pi} \Big(L\, \gamma^s_0(x) +
  c_1(x) \Big)+ \left( \frac{\als(\mu)}{4\pi}\right)^2 \bigg[ L^2 \left(\frac{1}{2} \Big(\gamma^s_0(x)\Big)^2 -
\beta_0 \gamma^s_0(x)\right)
\nn
\\
&& +
L \left(c_1(x)\Big(\gamma^s_0(x) - 2\beta_0\Big) 
+ \gamma^s_1(x)\right) + c_2(x) \bigg] 
+ \mathcal{O}\left(\als(\mu)^3\right).
\label{eq:rens}
\end{eqnarray}
The functions $c_1(x)$ and $c_2(x)$ must be determined by explicit perturbative calculations. The one-loop function, $c_1(x)$, is known \cite{1103.0550},
\begin{eqnarray}
c_1(x) &=& C_F \left[  \frac{1+x^2}{1-x^2} \left(-2 G^2(0;x)+8 G(1;x) G(0;x)-8 G(0,1;x)-\frac{4 \pi ^2}{3}\right)\nn
\right.\\&&\left. 
-4\frac{1+x}{1-x} G(0;x)\right],
\end{eqnarray}
but the two-loop function, $c_2(x)$, is one of our main new results and will be presented in Section \ref{sec:4}. Actually verifying that the renormalized
soft function has the form given in Eq.\ (\ref{eq:rens}) turns out to be a very useful check on the explicit two-loop calculation carried out in the next section.

\section{T\lowercase{he calculation of the soft function at one and two loops}}

The calculation of the $Q\bar{Q}$ soft function at two-loop order is non-trivial and deserves an in-depth discussion. As mentioned above, the leading order soft function was computed through the finite terms in reference~\cite{1103.0550}.
However, for our purposes, this is not enough; finite contributions from the charge and operator renormalization constants arise at the two-loop level which require the one-loop result to be known to
higher orders in $\ep$. Since it is relatively straightforward to compute, we give the one-loop result to all orders in $\ep$ in Section \ref{sec:oneloopsol}. The genuine two-loop contributions are significantly more
complicated and will be discussed at greater length. As explained in Section \ref{sec:2}, all purely virtual
corrections to the soft function vanish identically in dimensional regularization as a result of UV-IR cancellation. The
contributions to the soft function which must be considered at two-loop order can be classified as either real-virtual or real-real, depending on whether one or two soft partons appear in the final state.
Representative next-to-leading order diagrams are depicted in Figure 1.

\begin{figure}
\centering
\subfloat[]{\includegraphics[width=0.25\textwidth]{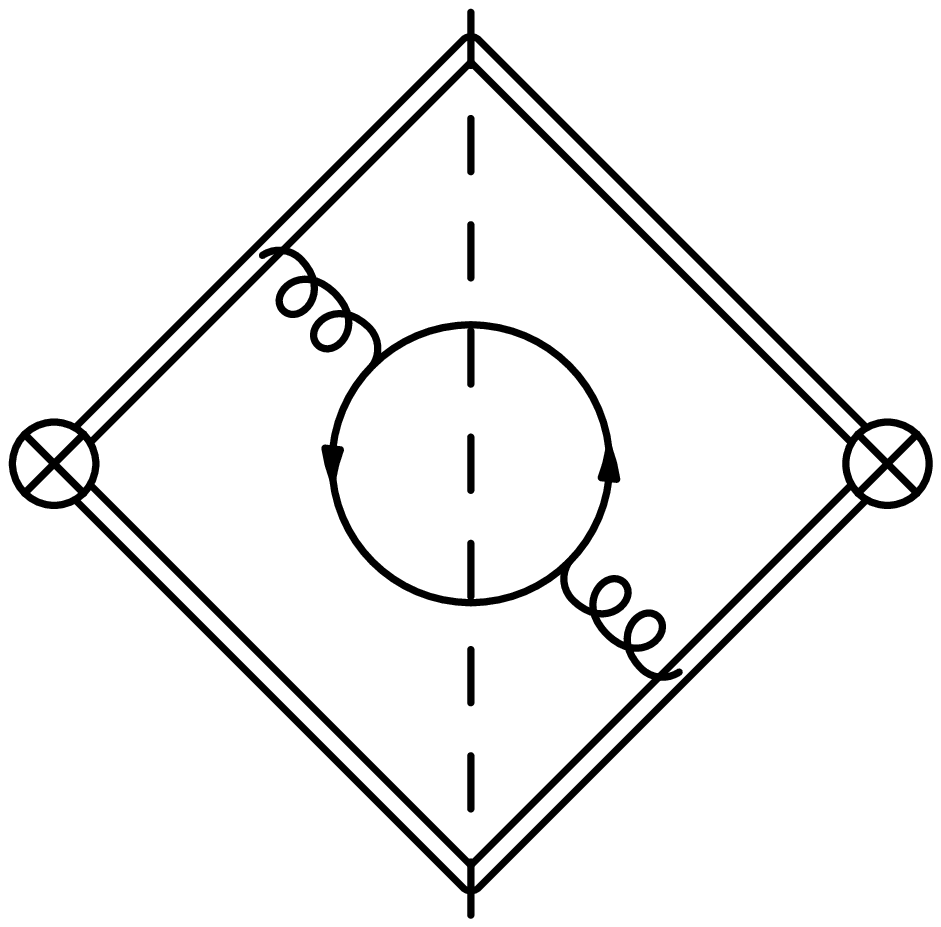}}
\qquad
\subfloat[]{\includegraphics[width=0.25\textwidth]{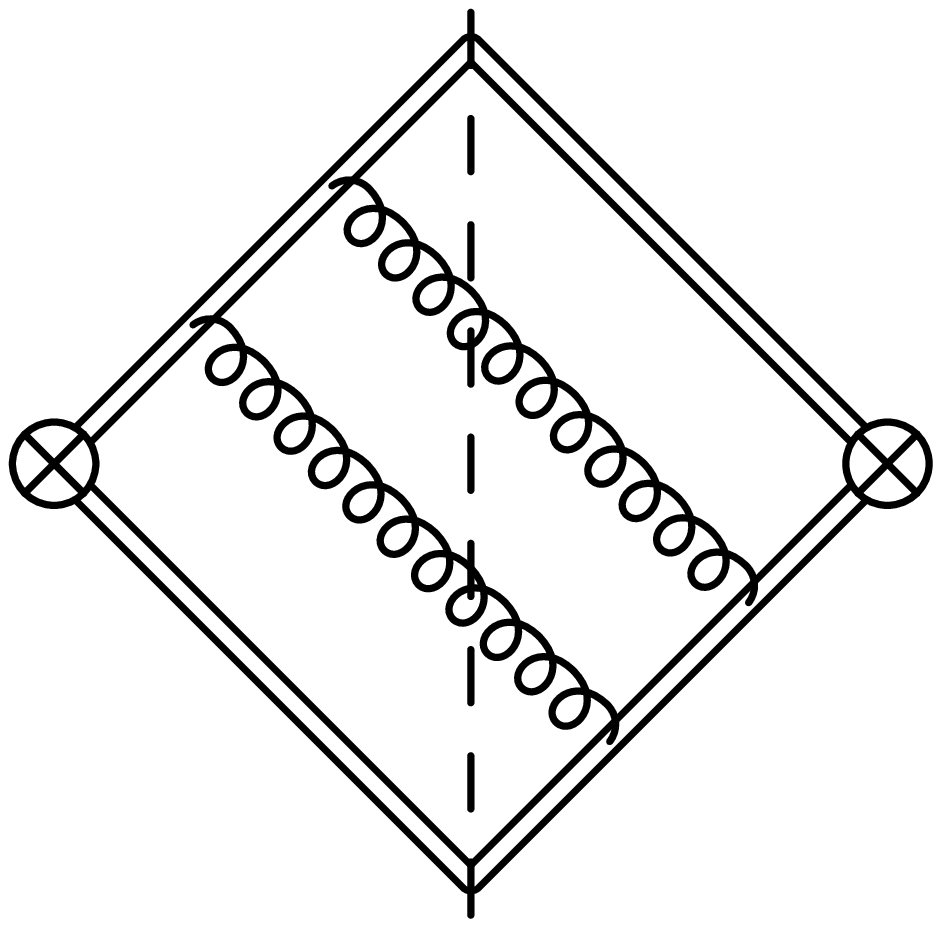}}
\\
\subfloat[]{\includegraphics[width=0.25\textwidth]{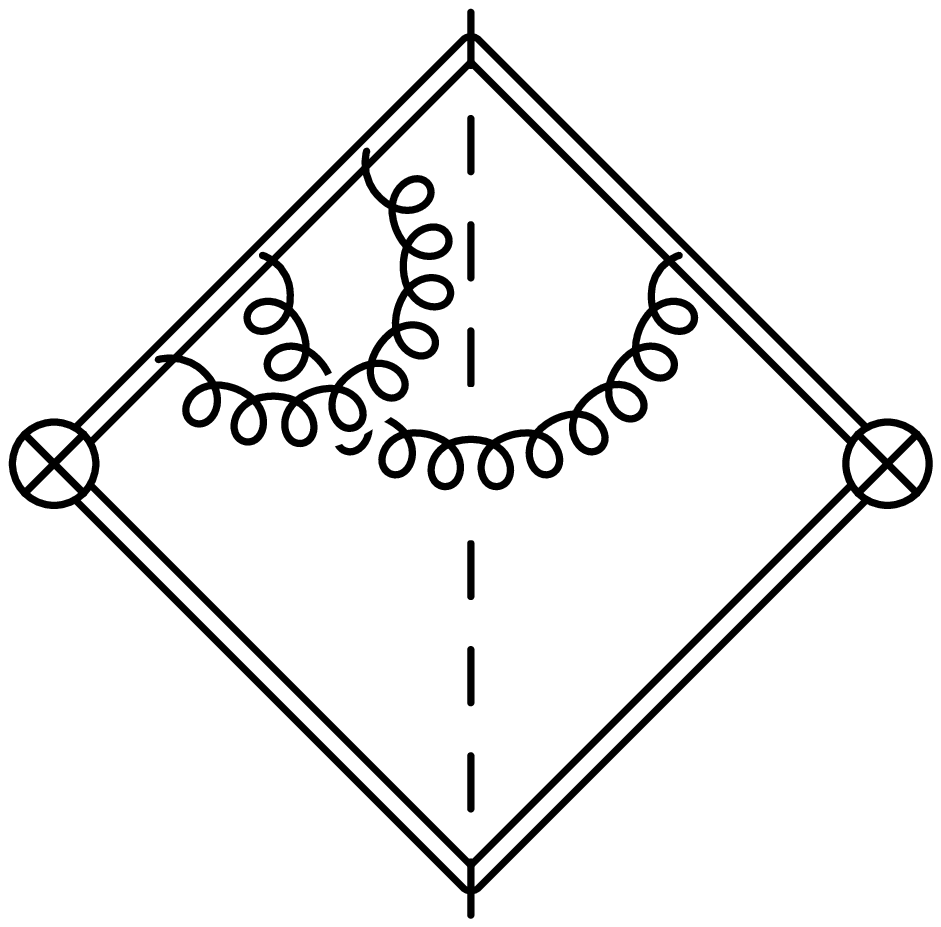}}
\qquad
\subfloat[]{\includegraphics[width=0.25\textwidth]{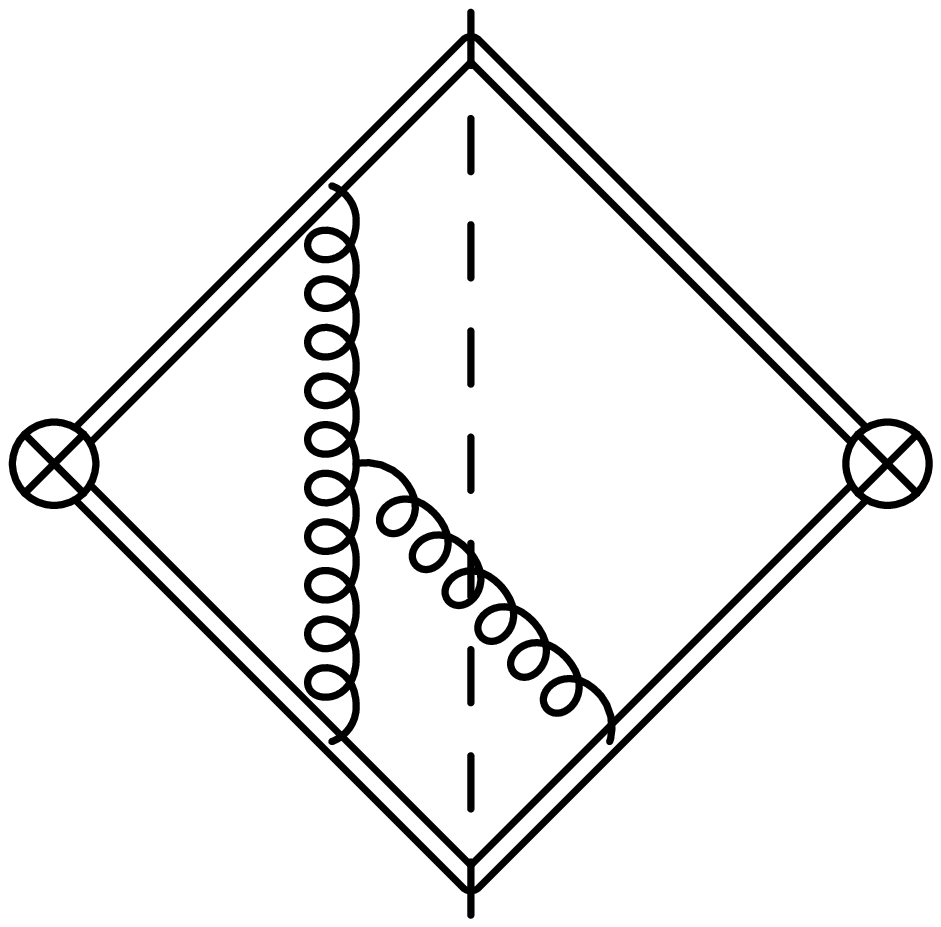}}
  \caption{Panels $(a)$ and $(b)$ show cut eikonal Feynman diagrams which contribute to $S_{Q\bar{Q}}^{{\rm RR}~(2)}(x,\ep)$ and panels $(c)$ and $(d)$
  show diagrams which contribute to $S_{Q\bar{Q}}^{{\rm RV}~(2)}(x,\ep)$.}
\label{fig:1}
\end{figure}
For convenience, we split the two-loop coefficient of the bare soft function, $S_{Q\bar{Q}}^{(2)}(x,\ep)$, into its real-virtual and real-real parts,
\begin{equation}
 S_{Q\bar{Q}}^{(2)}(x,\ep) = S_{Q\bar{Q}}^{{\rm RV}~(2)}(x,\ep) + S_{Q\bar{Q}}^{{\rm RR}~(2)}(x,\ep).
\end{equation}
The coefficients $S_{Q\bar{Q}}^{{\rm RV}~(2)}(x,\ep)$ and $S_{Q\bar{Q}}^{{\rm RR}~(2)}(x,\ep)$ will be treated in Sections \ref{sec:realvirtual} and \ref{sec:realreal} respectively.

\subsection{The one-loop soft function}
\label{sec:oneloopsol}

In this subsection, we solve the one-loop soft function once and for all. Due to the extreme simplicity of the integrand, there is no need to perform an integration by parts reduction~\cite{PHLTA.B100.65, NUPHA.B192.159};
a direct integration is straightforward to carry out and that is how we shall proceed. Evaluating the four cut eikonal diagrams that
contribute, we find
\begin{eqnarray}
\label{eq:oneloopsol}
S^{(1)}_{Q\bar{Q}}(x,\ep) &=& 32 \pi^3 (4\pi)^{-\ep} e^{\gamma_E  \epsilon }C_F \int \frac{\dd^d k}{(2\pi)^d} \delta\left(k^2\right)\delta \left(1 - (v + \bar{v})\cdot k\right)\times
\nn \\
&\times&\left( \frac{2 v \cdot \bar{v}}{v \cdot k ~\bar{v} \cdot k}
- \frac{v^2}{\left(v \cdot k\right)^2} - \frac{\bar{v}^2}{\left(\bar{v} \cdot k\right)^2}\right)
\nn \\&=&
   -\frac{4 e^{\gamma_E  \epsilon } C_F}{\Gamma (1-2 \epsilon )} \left(2 \Gamma (1-\epsilon )-(1+x) \left(\left(\frac{1-x}{1+x}\right)^2-2 \epsilon +1\right)\times\right.
\nn \\ 
   &\times&\left.\Big(\Gamma (-\epsilon ) \, _2F_1(1,1-\epsilon; 1 + \epsilon; x)+(1-x)^{2 \epsilon -1} x^{-\epsilon } \Gamma (1-2 \epsilon ) \Gamma (\epsilon )\Big)\vphantom{\left(\frac{(1-x)^2}{(1+x)^2}-2 \epsilon +1\right)}\right).
\end{eqnarray}

\subsection{The real-virtual contributions to the two-loop soft function}
\label{sec:realvirtual}

We generate the two-loop integrand via cut eikonal diagrams
and process its Lorentz and gauge structure in a semi-automated way.
For this purpose, we employ the programs {\tt QGRAF} \cite{JCTPA.105.279}, {\tt FeynCalc} \cite{NUPHZ.29A.204}, and {\tt ColorMath} \cite{1211.2099},
and find six independent, non-zero contributions to $S_{Q\bar{Q}}^{{\rm RV}~(2)}(x,\ep)$.
Ultimately, we arrive at an expression of the schematic form
\begin{equation}
S_{Q\bar{Q}}^{{\rm RV}~(2)}(x,\ep) = \int [dk \,dq] \mathcal{I}_{Q\bar{Q}}^{\rm RV}(q, k, x, \ep),
\end{equation}
where we have introduced the short-hand notation
\begin{equation}
\int [dk \,dq] = \int \dd^d k ~\Theta\left(k^0\right)\delta\left(k^2\right)  \delta\left(1 - \left(v+\bar{v}\right)\cdot k\right)\int \dd^d q\,.
\end{equation}
To simplify the calculation, we perform an integration by parts reduction using the development version of {\tt Reduze 2} \cite{1201.4330,Studerus:2009ye,cs.sc/0004015,fermat} which, among other new features,
allows for the application of Laporta's algorithm~\cite{hep-ph/0102033} to phase space integrals. 
The idea behind this was worked out some time ago and is commonly referred to as the reverse unitarity method~\cite{hep-ph/0207004,hep-ph/0306192,hep-ph/0312266}.
The key insight is that, for the purpose of integral reduction, one can simply replace delta function constraints with propagator denominators by virtue of the relation
\begin{equation}
\delta\left(k^2\right) = -\frac{1}{2\pi i} \left( \frac{1}{k^2+i0} - \frac{1}{ k^2 - i0} \right).
\end{equation}

After carrying out the integral reduction, we find
\begin{eqnarray}
\label{eq:resultugly}
&&S_{Q\bar{Q}}^{{\rm RV}~(2)}(x,\ep) = 2 C_A C_F {\rm Re}\left\{-\frac{4}{3 x (1+x)^4 \epsilon ^2 (1 + 2 \epsilon)^2} \left(-6 \left(x^3+x^2+1+x\right)^2+12 (1+x)^2 \times\nn
\right.\right. \\ &&\left.\left.
\times\left(45 x^4+28 x^3+30 x^2+28 x+45\right) \epsilon ^4-2
   \left(243 x^6+718 x^5+985 x^4+1052 x^3+985 x^2\nn
\right.\right.\right. \\ &&\left.\left.\left.
   +718 x+243\right) \epsilon ^3+\left(39 x^6+214 x^5+477 x^4+620 x^3+477 x^2+214 x+39\right) \epsilon ^2+\left(39 x^6\nn
\right.\right.\right. \\ &&\left.\left.\left.   
   +94 x^5+89 x^4+92 x^3+89
   x^2+94 x+39\right) \epsilon \vphantom{-6 \left(x^3+x^2+1+x\right)^2}\right)I^{\rm RV}_1(x,\ep)
   +\frac{32 (3-4 \epsilon)}{3 (1+x)^4 \epsilon^2 (1+2 \epsilon)^2}\times\nn
\right.\\ &&\left.  
   \times\left(\vphantom{-6 \left(x^3+x^2+1+x\right)^2}12 (1+x)^2 \left(5 x^2+2 x+5\right) \epsilon ^3+(1+x)^2 \left(1+x^2\right)-2 \left(7 x^4+28 x^3+38 x^2+28 x+7\right) \epsilon ^2\nn
\right.\right. \\ &&\left.\left.   
   -\left(5 x^4+12 x^3+10 x^2+12 x+5\right) \epsilon \vphantom{-6 \left(x^3+x^2+1+x\right)^2}\right)I^{\rm RV}_2(x,\ep)
   -\frac{8 x \left(1+x^2\right) (1 - 2 \epsilon)}{(1+x)^4 \epsilon  (1 + 2 \epsilon)} I^{\rm RV}_3(x,\ep) \nn
\right. \\ &&\left.     
   -\frac{4 }{x (1+x)^2 \epsilon  (1 + 2 \epsilon)} \left(-\left(1+x^2\right)^2+(2 x (x (x (3 x-10)-30)-10)+6) \epsilon ^2\nn
\right.\right. \\ &&\left.\left.    
   +(x (x (x (x+8)+26)+8)+1) \epsilon \vphantom{-\left(1+x^2\right)^2}\right) I^{\rm RV}_4(x,\ep)\nn
\right. \\ &&\left.   
   -\frac{8 (1- 2 \epsilon) \left(\left(1+x^2\right)^2+2 (x (x ((x-6) x-10)-6)+1) \epsilon \right)}{x (1+x)^2 \epsilon  (1 + 2\epsilon)} I^{\rm RV}_5(x,\ep) \nn
\right. \\ &&\left.   
-\frac{8 \left(1+x^4+x^2 (8 \epsilon +2)\right)}{(1+x)^4 (1+2 \epsilon)} I^{\rm RV}_6(x,\ep)
-\frac{4 \left(1+x^2+2 ((x-4) x+1) \epsilon\right)}{(1+x)^2 (1+2 \epsilon)} I^{\rm RV}_7(x,\ep) \right\},
\end{eqnarray}
where $\{I^{\rm RV}_1(x,\ep),\dots, I^{\rm RV}_7(x,\ep)\}$ are seven master integrals which must be calculated. Explicitly, we have
\begin{eqnarray}
I^{\rm RV}_1(x,\ep) &=& -i e^{2 \gamma_E  \epsilon } \pi ^{2 \epsilon -3}\int[dk\, dq] \frac{1}{\left(q\cdot \bar{v} + i 0\right) \, \left((k-q)^2+i 0\right)}
\\
I^{\rm RV}_2(x,\ep) &=& -i e^{2 \gamma_E  \epsilon } \pi ^{2 \epsilon -3}\int[dk\, dq] \frac{q^2}{\left(q\cdot \bar{v} + i 0\right) \,\left((k-q)^2 +  i 0\right)}
\nn
\\
I^{\rm RV}_3(x,\ep) &=& -i e^{2 \gamma_E  \epsilon } \pi ^{2 \epsilon -3}\int[dk\, dq] \frac{1}{k \cdot v\,\left(q\cdot \bar{v} +  i 0\right) \, \left((k-q)^2 + i 0\right)}
\nn
\\
I^{\rm RV}_4(x,\ep) &=& -i e^{2 \gamma_E  \epsilon } \pi ^{2 \epsilon -3}\int[dk\, dq] \frac{1}{\left(q^2 + i 0\right)\, \left(q\cdot \bar{v} + i 0\right) \, \left((k-q)\cdot v + i 0\right)}
\nn
\\
I^{\rm RV}_5(x,\ep) &=& -i e^{2 \gamma_E  \epsilon } \pi ^{2 \epsilon -3}\int[dk\, dq] \frac{k\cdot v}{\left(q^2 + i 0\right)\, \left(q\cdot \bar{v} + i 0\right) \, \left((k-q)\cdot v + i 0\right)}
\nn
\\
I^{\rm RV}_6(x,\ep) &=& -i e^{2 \gamma_E  \epsilon } \pi ^{2 \epsilon -3}\int[dk\, dq] \frac{1}{k\cdot v\, \left(q\cdot \bar{v} + i 0\right) \,\left( (k-q)\cdot v + i 0\right)\, \left( (k - q)^2 + i 0\right)}
\nn
\\
I^{\rm RV}_7(x,\ep) &=& -i e^{2 \gamma_E  \epsilon } \pi ^{2 \epsilon -3}\int[dk\, dq] \frac{1}{\left(q^2 + i 0\right)\, \left(q\cdot \bar{v} + i 0\right) \,\left( (k-q)\cdot v + i 0\right)\, \left( (k -q)^2 + i 0\right)}.\nn
\end{eqnarray}

It is worth pointing out that, in all cases, it is possible to integrate the master integrals
directly to sufficiently high order in $\ep$ by expanding in plus distributions under the integral sign. The disadvantage of this approach is that one would need to revisit the calculation at some point in the future in order to compute, say,
the three-loop soft function. Ideally, one would like to know the exact functional dependence on both $x$ and $\ep$. We were not able to find closed, all-orders-in-$\ep$ expressions in all cases so, instead, we employed a modern variant
of the method of differential equations \cite{PHLTA.B254.158,PHLTA.B259.314,PHLTA.B267.123,hep-ph/9212308,hep-ph/9306240,hep-th/9711188,Gehrmann1}.
Specifically, we decouple our system of differential equations to all orders in $\ep$ by constructing a {\it normal form basis} for the system~\cite{Kotikov:2010gf,1304.1806}.
In a nutshell, the idea is to work in a new basis built out of master integrals which take the form of so-called pure functions.
To fix the integration constants, we specify simple regularity conditions in the physical region.
For integrals unconstrained by these conditions, we provide closed form expressions which are exact in both $x$ and $\ep$.
This setup allows for a straightforward and transparent calculation of the master integrals to any desired order in the epsilon expansion.
To the best of our knowledge, our calculation is the first application of the normal form basis method to the calculation of {\it eikonal phase space integrals}
(see the recent reference~\cite{Hoschele:2014qsa} for an application to standard phase space integrals).
Here, we are interested in the real parts of the master integrals, which we calculate directly in the physical region of phase space, $0<x<1$. To illustrate the power of our method and to check an unexpected analytical property
of our result (see Section \ref{sec:4} below), we actually calculate our master integrals to one order higher in $\ep$ than is strictly necessary for a two-loop calculation of the soft function. As a result, we shall actually obtain
and present results for $S_{Q\bar{Q}}^{{\rm RV}~(2)}(x,\ep)$ up to and including terms of $\mathcal{O}\left(\ep^2\right)$.

Let us begin by briefly reviewing the proposal of reference \cite{1304.1806} in the context of the present calculation.
One begins with a first-order system of differential equations for ${\bf I}^{\rm RV}(x,\ep) = \{I^{\rm RV}_1(x,\ep),\dots, I^{\rm RV}_7(x,\ep)\}$ derived using {\tt Reduze 2} or some other reduction program,
\begin{equation}
\label{eq:initRVsystem}
\frac{\dd}{\dd x}{\bf I}^{\rm RV}(x, \ep) = \underaccent{\tilde}{{\bf A}}^{\rm RV}(x, \ep){\bf I}^{\rm RV}(x, \ep).
\end{equation}
The idea is to then attempt to find a new basis,
\begin{equation}
{\bf I}^{\rm RV}(x, \ep) = \underaccent{\tilde}{{\bf B}}^{\rm RV}(x, \ep){\bf F}^{\rm RV}(x, \ep),
\end{equation}
where $F^{\rm RV}_i(x, \epsilon) = \sum_{n = 0}^\infty c_i^{(n)}(x) \epsilon^n$ for some coefficient functions, $c_i^{(n)}(x)$, of uniform transcendentality weight $n$.
Once this has been achieved, the linear map $\underaccent{\tilde}{{\bf B}}^{\rm RV}(x, \ep)$ transforms Eq.\ (\ref{eq:initRVsystem})
into a system of differential equations of the form
\begin{equation}
\label{eq:RVHennsystem}
\frac{\dd}{\dd x}{\bf F}^{\rm RV}(x, \ep) = \ep\, \underaccent{\tilde}{{\bf H}}^{\rm RV}(x) {\bf F}^{\rm RV}(x, \ep),
\end{equation}
which implies complete decoupling after expansion in $\ep$.
${\bf F}^{\rm RV}(x, \ep)$ is what we refer to in this work as a normal form basis. 
Elements of a normal form basis, in this case the $F^{\rm RV}_i(x, \epsilon)$ introduced above, are usually referred to as pure functions in the literature; 
due to the fact that the terms in their $\epsilon$ expansions are of uniform transcendentality weight, starting at weight zero at $\mathcal{O}\left(\epsilon^0\right)$,
they can be thought of formally as weight zero objects if one assigns a transcendentality weight of $-1$ to $\epsilon$.
Once appropriate input integrals have been evaluated, Eq.\ (\ref{eq:RVHennsystem}) is significantly easier
to solve than Eq.\ (\ref{eq:initRVsystem}) because the coefficient functions $c_i^{(n)}(x)$ are completely determined by the coefficient functions $c_i^{(n - 1)}(x)$ up to constants.

Unfortunately, there is no fully-general, systematic approach to the construction of a normal form basis, Eq.\ (\ref{eq:RVHennsystem}).
However, hints for finding normal form integrals have been given \cite{1304.1806,1306.2799,1401.2979} and,
more recently, a recipe to pass from the form Eq.\ (\ref{eq:initRVsystem}) to the form Eq.\ (\ref{eq:RVHennsystem}) which works
in many cases of practical interest was described in \cite{1404.4853}.
While these ideas would be useful in the present context, the bulk of this work
was carried out before reference \cite{1404.4853} appeared, and we use a somewhat more
{\it ad hoc} method to find the matrix $\underaccent{\tilde}{{\bf B}}^{\rm RV}(x, \ep)$ here.
It turns out that the integrals $I^{\rm RV}_6(x,\ep)$ and $I^{\rm RV}_7(x,\ep)$ are, up to normalization, already pure functions.
In several cases, normal form basis elements could be constructed by squaring a propagator and then adjusting the overall normalization.
The rest of the integrals in our normal form basis are more non-trivial and required us to consider linear combinations of several integrals with propagators raised to higher powers.
Here, we resorted to an explicit calculation of all of the integrals up to transcendentality weight three.
We then made ans\"{a}tze for the undetermined coefficients in the basis change.
The undetermined coefficients in our ans\"{a}tze happened to be tightly constrained by the contributions of low transcendentality weight that we computed
explicitly and this allowed us to determine a complete basis of pure functions in an experimental fashion.

We now present the results that we obtained by going through the steps described above. By making the change of variables
\begin{eqnarray}
\label{eq:RVvarchange}
I^{\rm RV}_1(x,\ep) &=& \frac{(1+x) \left(x^2 (3 \epsilon -1)-2 x \epsilon +3 \epsilon -1\right)}{4 (1-x) x (1-2 \epsilon )^2 \epsilon  (4 \epsilon -1)}
F_1^{\rm RV}(x, \ep) \nn
-\frac{(1+x)^2 (3 \epsilon -1)}{16 x (1-2 \epsilon )^2 \epsilon  (4 \epsilon -1)}
F_2^{\rm RV}(x, \ep)
\nn
\\
I^{\rm RV}_2(x,\ep) &=& -\frac{(1+x)}{32 x^2 (1-x) \epsilon  (4 \epsilon -3) (4 \epsilon -1) (1-2 \epsilon)^2}
\Big(6 x^4+(x (x (3 x (9 x-4)-14)-12)
\nn
\\
&&
+27) \epsilon ^2+(x ((8-27 x) x+6)+8) x \epsilon
 -27 \epsilon +6\Big)F_1^{\rm RV}(x, \ep)
\nn
\\
&&
 +\frac{(1+x)^2 (3 \epsilon -2) \left(x^2 (9 \epsilon -3)+2 x \epsilon +9 \epsilon -3\right)}{128 x^2 \epsilon  (4 \epsilon -3) (4 \epsilon -1) (1-2 \epsilon )^2} F_2^{\rm RV}(x, \ep) 
\nn
\\
I^{\rm RV}_3(x,\ep) &=& -\frac{(1+x)^3}{2 (1-x) x \epsilon  \left(1-2\epsilon\right)\left(1-4\epsilon\right)}F_1^{\rm RV}(x, \ep) 
+\frac{(1+x)^2}{8 x \epsilon  \left(1-2\epsilon\right)\left(1-4\epsilon\right)}F_2^{\rm RV}(x, \ep)
\nn
\\
&&
+\frac{(1+x)^3}{4 (1-x) x \epsilon ^2 (2 \epsilon -1)}F_3^{\rm RV}(x, \ep) 
\nn
\\
I^{\rm RV}_4(x,\ep) &=& -\frac{(1+x)^2}{2 (1 - x)^2 \epsilon ^2 (1-4 \epsilon)}F_4^{\rm RV}(x, \ep) 
+\frac{(1+x)}{4 (1-x) \epsilon ^2 (1-4 \epsilon)}F_5^{\rm RV}(x, \ep) 
\nn
\\
I^{\rm RV}_5(x,\ep) &=& \frac{\left(x^2 (3 \epsilon -1)-2 x \epsilon +3 \epsilon -1\right)}{4 (1 - x)^2 \epsilon ^2 \left(1-2\epsilon\right)\left(1-4\epsilon\right)}F_4^{\rm RV}(x, \ep) 
-\frac{(1+x) (3 \epsilon -1)}{8 (1-x) \epsilon ^2 \left(1-2\epsilon\right)\left(1-4\epsilon\right)}F_5^{\rm RV}(x, \ep) 
\nn
\\
I^{\rm RV}_6(x,\ep) &=& \frac{ (1+x)^2}{(1-x)^2 \epsilon ^3}F_6^{\rm RV}(x, \ep)
\nn
\\
I^{\rm RV}_7(x,\ep) &=& \frac{(1+x)}{(1 - x) \epsilon ^3}F_7^{\rm RV}(x, \ep) ,
\end{eqnarray}
where the $F_i^{\rm RV}(x, \ep)$ satisfy simple differential equations of the form
\begin{eqnarray}
\label{eq:RVnormalform}
\frac{\dd}{\dd x} F_1^{\rm RV}(x, \ep) &=& \ep\left(- \left(\frac{1}{x}+\frac{2}{1-x}\right)F_1^{\rm RV}(x, \ep)-\frac{1}{4 x}F_2^{\rm RV}(x, \ep)\right)
\nn
\\
\frac{\dd}{\dd x} F_2^{\rm RV}(x, \ep) &=&-\frac{8 \ep}{x}F_1^{\rm RV}(x,\ep)
\nn
\\
\frac{\dd}{\dd x} F_3^{\rm RV}(x, \ep) &=&\ep\left(2 \left(\frac{1}{x}-\frac{2}{1+x}\right) F_1^{\rm RV}(x, \ep)-\frac{1}{2 x}F_2^{\rm RV}(x, \ep)\right.
\nn
\\
&&
\left. -2\left(\frac{1}{1+x}+\frac{1}{1-x}\right) F_3^{\rm RV}(x, \ep)\right)
\nn
\\
\frac{\dd}{\dd x} F_4^{\rm RV}(x, \ep) &=&\ep\left(\frac{2 }{x}F_1^{\rm RV}(x, \ep)- \left(\frac{3}{x}-\frac{2}{1+x}+\frac{4}{1-x}\right)F_4^{\rm RV}(x, \ep)-\frac{1}{2 x}F_5^{\rm RV}(x, \ep)\right)
\nn
\\
\frac{\dd}{\dd x} F_5^{\rm RV}(x, \ep) &=&\ep\left(\frac{1}{x}F_2^{\rm RV}(x, \ep)-\frac{4}{x}F_4^{\rm RV}(x, \ep)-2 
   \left(\frac{1}{x}-\frac{1}{1+x}+\frac{1}{1-x}\right)F_5^{\rm RV}(x, \ep)\right)
\nn
\\
\frac{\dd}{\dd x} F_6^{\rm RV}(x, \ep) &=&\ep\left(\frac{1}{2 x}F_3^{\rm RV}(x, \ep)+ \left(\frac{1}{2 x}-\frac{1}{1+x}\right)F_4^{\rm RV}(x, \ep)-\frac{1}{4 x}F_5^{\rm RV}(x, \ep)\right.
\nn
\\
&&
\left. -2  \left(\frac{1}{x}+\frac{2}{1-x}\right)F_6^{\rm RV}(x, \ep)\right)
\nn
\\
\frac{\dd}{\dd x} F_7^{\rm RV}(x, \ep) &=&\ep\left(\frac{1}{2 x}F_2^{\rm RV}(x, \ep)+\frac{4}{x}F_4^{\rm RV}(x, \ep)-2 \left(\frac{1}{1+x}+\frac{1}{1-x}\right) F_7^{\rm RV}(x, \ep)\right),
\end{eqnarray}
we arrive at
\begin{eqnarray}
\label{eq:resultclean}
&&S_{Q\bar{Q}}^{{\rm RV}~(2)}(x,\ep) = 2 C_A C_F {\rm Re}\left\{\frac{4 \left(2 \left(1+x^2\right) \epsilon +x^2-8 (1+x)^2 \epsilon ^2+1\right)}{\left(1-x^2\right) \epsilon ^3 (2 \epsilon +1)^2}F_1^{\rm RV}(x, \ep)
\right.
\nn
\\
&&
\left.
-\frac{4 (3 \epsilon+1)}{\epsilon ^2 (2 \epsilon +1)^2} F_2^{\rm RV}(x, \ep) 
+\frac{2 \left(1+x^2\right)}{\left(1-x^2\right) \epsilon ^3 (2 \epsilon +1)} F_3^{\rm RV}(x, \ep) 
-\frac{4 }{\left(1-x^2\right)^2 \epsilon ^3 (2 \epsilon +1)} \left(\left(1+x^2\right)^2
\right.\right.
\nn
\\
&&
\left.\left.
+4 (x (x ((x-2)x-4)-2)+1) \epsilon \vphantom{\left(1+x^2\right)^2}\right) F_4^{\rm RV}(x, \ep)
+\frac{4 \left(1+x^2\right)}{\left(1-x^2\right) \epsilon ^2 (2 \epsilon +1)} F_5^{\rm RV}(x, \ep) 
\right.
\nn
\\
&&
\left.
-\frac{8 \left(x^4+x^2 (8 \epsilon +2)+1\right)}{\left(1-x^2\right)^2 \epsilon ^3 (2 \epsilon +1)} F_6^{\rm RV}(x, \ep)
-\frac{4 \left(x^2+2 ((x-4) x+1) \epsilon +1\right)}{\left(1-x^2\right) \epsilon ^3 (2\epsilon +1)} F_7^{\rm RV}(x, \ep) \right\}.
\end{eqnarray}
The above expression for $S_{Q\bar{Q}}^{{\rm RV}~(2)}(x,\ep)$ is cleaner and more transparent than that given previously in Eq.\ (\ref{eq:resultugly}). Not only is Eq.\ (\ref{eq:resultclean})
somewhat more compact, but we can clearly see from it that the reduced integrand obtained by {\tt Reduze 2} out of the box obscures the fact that there are no power-law divergences in our result at $x = 0$.

Most integration constants can be fixed by so-called regularity conditions. The key idea is relatively simple, although determining the complete set of useful constraints for a given problem
requires careful analysis. As is clear from the form of Eqs.\ (\ref{eq:RVnormalform}), the only potential singularities of the differential equations lie at the points $x = -1$, $x = 0$, and $x = 1$.
Many of these singularities, however, are actually absent in particular master integrals,
as one can see, for example, by examining their cut structure.
When a master integral is known to be regular or vanishing in a particular limit, one can employ this information to determine integration constants
such that the expressions obtained respect these constraints.
It should be stressed that, at the technical level, this is not always an easy program to carry out,
especially in more complicated situations with more scales where one obtains a first-order system of partial differential equations. 
It turns out that three of our seven master integrals, $I_1^{\rm RV}(x, \ep)$, $I_3^{\rm RV}(x, \ep)$,
and $I_7^{\rm RV}(x, \ep)$, are completely constrained by a regularity condition at $x = 1$ (the threshold limit).
Further constraints could be obtained by considering the master integrals at the point $x = -1$.
Here, we chose to work exclusively in the physical region $0<x<1$ and do not
consider regularity constraints at $x=-1$ in order to avoid performing explicit analytical
continuations to the unphysical region, $x<0$.
Instead, we fix the remaining integration constants by direct calculation.

By direct integration, we obtained closed form expressions, exact in both $x$ and $\ep$,
for the following integrals,\footnote{The special function $F_1(a; b_1, b_2; c; x, y)$
appearing in Eqs.\ (\ref{eq:RVexplicit}) is called Appell's $F_1$ function and has the standard definition
\begin{equation}
F_1(a; b_1, b_2; c; x, y) = \frac{\Gamma(c)}{\Gamma(a)\Gamma(b_1)\Gamma(b_2)}\sum_{n = 0}^\infty \sum_{m = 0}^\infty \frac{\Gamma(a + n + m)\Gamma(b_1 + n)\Gamma(b_2 + m)}{\Gamma(c + n + m)}\frac{x^n}{n!}\frac{y^m}{m!}.
\end{equation}
}
\begin{eqnarray}
\label{eq:RVexplicit}
I_2^{\rm RV}(x, \ep) &=&  -\frac{1}{2}e^{2 i \pi  \epsilon } e^{2 \gamma_E  \epsilon} (1+x) x^{-2 + \epsilon} \Gamma (2-\epsilon) \Gamma (2 \epsilon -2)\times
\nn
\\
&&
\times\left(\frac{x^{4-3 \epsilon } \Gamma (3 \epsilon -4) \, _2F_1(5-4 \epsilon ,1-\epsilon
   ;5-3 \epsilon ;x)}{\Gamma (2 \epsilon -3)}\right.
\nn
\\
&&
\left.+\frac{\Gamma (4-3 \epsilon ) \, _2F_1(1-\epsilon ,2 \epsilon -3;3 \epsilon - 3;x)}{\Gamma (5-4 \epsilon )}\right)
\nn
\\
I_4^{\rm RV}(x, \ep) &=& -\frac{e^{2 i \pi  \epsilon } e^{2 \gamma_E  \epsilon} \Gamma (2 \epsilon )}{2 \Gamma (1+\epsilon)} \Big((1+x)^2 x^{\epsilon } \Gamma (1-\epsilon ) \Gamma (\epsilon )
\, _2F_1\left(1,1-\epsilon ;1+\epsilon;x^2\right)
\nn
\\
&&
-2 (1-x)^{2 \epsilon -1} x^{-\epsilon } (1+x)^{1+2 \epsilon} \Gamma (-2 \epsilon ) \Gamma^2 (1+\epsilon)\Big)\times
\nn
\\
&&
\times\left(\frac{x^{1-3 \epsilon } \Gamma (3 \epsilon -1) \, _2F_1(2-4 \epsilon ,1-\epsilon ;2-3\epsilon ;x)}{\Gamma (2 \epsilon )}\right.
\nn
\\
&&
\left.+\frac{\Gamma (1-3 \epsilon )\, _2F_1(1-\epsilon ,2 \epsilon ;3 \epsilon ;x)}{\Gamma (2-4 \epsilon )}\right)+2 e^{i \pi  \epsilon }e^{2 \gamma_E  \epsilon}\cos (\pi  \epsilon )
   x^{-\epsilon } (1+x)^{2 \epsilon +1}  \times
\nn
\\
&&
\times   \Gamma (-2 \epsilon )\Gamma (\epsilon ) \Gamma (2 \epsilon ) \Big(-(1-\epsilon) \, _2F_1(1-2 \epsilon ,-\epsilon ;1-\epsilon;x)+(1-2 \epsilon )\times
\nn
\\
&&
\times \, _2F_1(2-2 \epsilon ,-\epsilon ;1-\epsilon ;x)\Big) \left(\frac{x^{1-3 \epsilon } \Gamma (3 \epsilon -1) \, _2F_1(2-4 \epsilon ,1-\epsilon ;2-3 \epsilon ;x)}{\Gamma (2 \epsilon)}
\right.   
\nn
\\
&&
\left.   +\frac{\Gamma (1-3 \epsilon ) \, _2F_1(1-\epsilon ,2 \epsilon ;3 \epsilon ;x)}{\Gamma (2-4 \epsilon )}\right)
\nn   
\\
I_5^{\rm RV}(x, \ep) &=& -\frac{e^{2 i \pi  \epsilon} e^{2 \gamma_E  \epsilon} \Gamma (2 \epsilon )}{2 \Gamma (3-4 \epsilon ) \Gamma (1+\epsilon) \Gamma (2
   \epsilon -1)} \Big((1+x) x^{\epsilon } \Gamma (1-\epsilon ) \Gamma (\epsilon ) \, _2F_1\left(1,1-\epsilon ;1+\epsilon;x^2\right)
\nn
\\
&&
   -2 (1-x)^{2 \epsilon -1} x^{-\epsilon } (1+x)^{2 \epsilon } \Gamma (-2 \epsilon ) \Gamma^2 (1+\epsilon)\Big) \times
\nn
\\
&&
   \times\Big(x^{2-3 \epsilon } \Gamma (3-4 \epsilon ) \Gamma (3 \epsilon -2)\,_2F_1(3-4 \epsilon,1-\epsilon ;3-3 \epsilon ;x)
\nn
\\
&&   
+\Gamma (2-3 \epsilon ) \Gamma (2 \epsilon -1) \, _2F_1(1-\epsilon ,2 \epsilon -1;3 \epsilon -1;x)\Big)
\nn
\\
&&
+\frac{2 e^{i \pi  \epsilon }e^{2 \gamma_E  \epsilon }
\cos (\pi  \epsilon ) x^{-\epsilon } (1+x)^{2 \epsilon }  \Gamma (-2 \epsilon ) \Gamma (\epsilon ) \Gamma (2 \epsilon )}{\Gamma (3-4 \epsilon ) \Gamma (2 \epsilon-1)}\times
\nn
\\
&&
 \times \Big(-(1-\epsilon)\, _2F_1(1-2 \epsilon ,-\epsilon ;1-\epsilon ;x)+(1-2 \epsilon ) \, _2F_1(2-2 \epsilon ,-\epsilon ;1-\epsilon ;x)\Big)\times
\nn
\\
&&
 \times\Big(x^{2-3 \epsilon } \Gamma (3-4 \epsilon ) \Gamma (3 \epsilon -2) \,_2F_1(3-4 \epsilon ,1-\epsilon ;3-3 \epsilon ;x)
\nn
\\
&&
+\Gamma (2-3 \epsilon ) \Gamma (2 \epsilon -1) \, _2F_1(1-\epsilon ,2 \epsilon -1;3 \epsilon -1;x)\Big)
\nn
\\
I_6^{\rm RV}(x, \ep) &=&  \frac{e^{2 i \pi \epsilon} e^{2\gamma_E \epsilon}\Gamma(1-\epsilon)\Gamma(2\epsilon)}{2 \Gamma(2-2\epsilon)\Gamma(1+\epsilon)}F_1\left(1-\epsilon; 2\epsilon, 1; 2 - 2\epsilon; 1-x, \frac{x-1}{x}\right)\times
\nn\\&&
\times\Big(2 x^{-1-\epsilon}(1-x)^{-1+2\epsilon}(1+x)^{2+2\epsilon}\Gamma(-2\epsilon)\Gamma^2(1+\epsilon)
\nn\\&&
- x^{-1+\epsilon}(1+x)^3\Gamma(1-\epsilon) \Gamma(\epsilon) _2F_1\left(1, 1-\epsilon; 1+\epsilon;x^2\right)\Big) + 2 e^{i \pi \epsilon}e^{2\gamma_E \epsilon}\cos(\pi \epsilon)\times
\nn\\&&
\times \frac{\Gamma(1-\epsilon)\Gamma(-2\epsilon)\Gamma(\epsilon)\Gamma(2\epsilon)}{\Gamma(2-2\epsilon)} x^{-1-\epsilon} (1 + x)^{2+2\epsilon}\times
\nn\\&&
\times  F_1\left(1-\epsilon; 2\epsilon, 1; 2-2\epsilon; 1-x, \frac{x-1}{x}\right)\Big((1-2\epsilon)_2F_1\left(2-2\epsilon, -\epsilon; 1-\epsilon; x\right)
\nn\\&&
- (1-\epsilon) _2F_1\left(1-2\epsilon, -\epsilon; 1-\epsilon; x\right) \Big).
\end{eqnarray}
These explicit solutions can be used to fix all remaining integration constants.\footnote{Correctly computing the phases that appear in Eqs.\ (\ref{eq:RVexplicit}) is, in our opinion, one of the more challenging aspects of the calculation.
We refer the interested reader to reference \cite{1107.4384} where many details are given for a very similar soft, real-virtual, two-loop calculation in the presence of heavy quarks.}
While for that purpose it would actually be sufficient to have a result in some
asymptotic limit, the full $x$ dependence given above allows for a check of the
differential equations themselves.
In order to extract the required integration constants, we
expanded the Gauss hypergeometric functions using the {\tt Mathematica} program {\tt HypExp} \cite{hep-ph/0507094, 0708.2443}
and the Appell $F_1$ functions using Picard's integral,
\begin{equation}
F_1(a; b_1, b_2; c; x, y) = \frac{\Gamma(c)}{\Gamma(a)\Gamma(c-a)}\int_0^1 \dd t\, t^{a-1} (1-t)^{c-a-1} (1 - x t)^{-b_1} (1 - y t)^{-b_2}.
\end{equation}

With the presented ingredients it is straightforward to integrate the system
of differential equations for the real-virtual contributions to arbitrarily high
orders in the dimensional regularization parameter.
We refrain from explicitly writing our result
for $S_{Q\bar{Q}}^{{\rm RV}~(2)}(x,\ep)$ to $\mathcal{O}\left(\ep^2\right)$ accuracy because of its length,
but we have included it with our submission to arXiv.org as an ancillary file.

\subsection{The real-real contributions to the two-loop soft function}
\label{sec:realreal}

The real-real corrections are actually easier to calculate than the real-virtual corrections because all $x$-dependent master integrals which contribute are completely determined by the differential equations they satisfy,
regularity conditions at threshold ($x = 1$), and a single, trivial input integral.
As before, we use a variety of scripts to evaluate the thirteen independent, non-zero contributions\footnote{We use Feynman gauge everywhere in the calculation. A different number of independent cut eikonal 
diagrams would result, for example, in lightcone gauge.}
to the square of the tree-level, two-emission amplitude and arrive at an expression of the form
\begin{equation}
 S_{Q\bar{Q}}^{{\rm RR}~(2)}(x,\ep) = \int [dk_1 \,dk_2] \mathcal{I}_{Q\bar{Q}}^{\rm RR}(k_1, k_2, x, \ep),
\end{equation}
where we have introduced the short-hand notation
\begin{equation}
\int [dk_1 \,dk_2] = \int \dd^d k_1 \dd^d k_2~\Theta\left(k_1^0\right)\Theta\left(k_2^0\right)\delta\left(k_1^2\right) \delta\left(k_2^2\right) \delta\left(1 - \left(v+\bar{v}\right)\cdot \left(k_1 + k_2\right)\right) 
\end{equation}
to save space. Once again, we perform an integration by parts reduction and find
\begin{eqnarray}
\label{eq:RRresultugly}
&&S_{Q\bar{Q}}^{{\rm RR}~(2)}(x,\ep) = C_F^2 \left[\vphantom{\frac{128 (1-4 \epsilon ) \left(1 + x^2 - (1+x)^2 \epsilon \right)^2}{(1+x)^4 \epsilon } I^{\rm RR}_4(x, \ep)}
-\frac{256 (1-2 \epsilon ) (1-4 \epsilon ) (3-4 \epsilon )}{\epsilon }I^{\rm RR}_1(x, \ep)
+ \frac{512 (1-2 \epsilon ) (1-4 \epsilon )}{(1+x)^2 \epsilon}\times
\right. \nn \\ && \left.
\times  \Big(1+x^2-(1+x)^2 \epsilon \Big)I^{\rm RR}_2(x, \ep)
-\frac{128 (1-4 \epsilon ) \left(1 + x^2 - (1+x)^2 \epsilon \right)^2}{(1+x)^4 \epsilon } I^{\rm RR}_4(x, \ep)\right]
\nn \\ &&
+ C_F n_f T_F \left[- \frac{128  (1-x)^2 (1-\epsilon )(1-2\epsilon )(3-4 \epsilon ) }{x (3-2 \epsilon ) \epsilon }I^{\rm RR}_1(x, \ep)
+ \frac{128 (1-\epsilon ) (1-2 \epsilon )}{x (1 + x)^2 (3-2 \epsilon ) \epsilon}\times
\right. \nn \\ && \left. 
\times   (1-x)^2 \Big(1 + x^2 - 2 (1+x)^2 \epsilon \Big)I^{\rm RR}_3(x, \ep)
\vphantom{\frac{128 I^{\rm RR}_1(x, \ep) (1-x)^2 (1-\epsilon )(1-2\epsilon )(3-4 \epsilon ) }{x (3-2 \epsilon ) \epsilon }}\right]
\nn \\ &&
+ C_A C_F \left[\frac{16 (3-4 \epsilon)}{(1+x)^2 x \epsilon ^2 (1+2 \epsilon )^2 (1 - 2 \epsilon) (3 - 2 \epsilon)} \Big(256 (1+x)^2 \left(1+x^2\right) \epsilon ^7
\right. \nn \\ && \left. 
-32 (1+x)^2(x (23 x-18)+23) \epsilon ^6+32 (x (x (x (17 x-16)-68)-16)+17) \epsilon ^5
\right. \nn \\ && \left. 
+32 (x (x (x (4 x+55)+107)+55)+4) \epsilon ^4-16 (x (x (x (13 x+72)+121)+72)+13) \epsilon ^3
\right. \nn \\ && \left. 
+2 (x (x (x(x+36)+34)+36)+1) \epsilon ^2+2 (1+x)^2 (x (7 x+46)+7) \epsilon 
\right. \nn \\ && \left. 
+3 (1+x)^2 ((x-8) x+1)\Big)I^{\rm RR}_1(x, \ep) 
-\frac{32  (1-2 \epsilon)}{(1+x)^2 \epsilon ^2 (1+2 \epsilon)^2} \Big(6 \left(1+x^2\right) \epsilon-x^2 
\right. \nn \\ && \left. 
+ 64 (1+x)^2 \epsilon ^5-16 (x-3) (3 x-1) \epsilon ^4-8 (x (11 x-4)+11) \epsilon ^3+8 (2+x) (1+2 x) \epsilon ^2
\right. \nn \\ && \left.
-1\Big)I^{\rm RR}_2(x, \ep)-\frac{16 }{(1+x)^4 x \epsilon ^2 (3 - 2 \epsilon) (1+2 \epsilon )}\Big(-16 \left(1-x^2\right)^2 (x (27 x+38)+27) \epsilon ^5
\right. \nn \\ && \left.
+8 \left(1-x^2\right)^2 \left(1+x^2\right) \epsilon+3 \left(1-x^2\right)^2 \left(1+x^2\right)+128 (1-x)^2 (1+x)^4 \epsilon ^6
\right. \nn \\ && \left.
+8 (x (x (x (x (x (61 x+36)-57)-72)-57)+36)+61) \epsilon ^4
\right. \nn \\ && \left.
-4 (x (x (x (x (x (45 x-38)-21)+76)-21)-38)+45) \epsilon ^3
\right. \nn \\ && \left.
-2 (x (x (x (x (x (7 x+44)-43)-88)-43)+44)+7) \epsilon ^2\Big)I^{\rm RR}_3(x, \ep) 
+\frac{8 (1-4 \epsilon)}{(1+x)^4 \epsilon  (1+2 \epsilon )^2} \times
\right. \nn \\ && \left.
\times \Big(-8 \left(1-x^2\right)^2 \epsilon ^3+5 \left(1+x^2\right)^2+16 (1+x)^4 \epsilon ^4-8 (x (x (x (x+10)-4)+10)+1) \epsilon ^2
\right. \nn \\ && \left.
+2 (1-x)^2 (x (7 x-2)+7) \epsilon \Big)I^{\rm RR}_4(x, \ep) 
-\frac{32  x \left(1 + x^2\right) (12 (\epsilon -1) \epsilon +1)}{(1+x)^4 \epsilon  (1+2 \epsilon )}I^{\rm RR}_5(x, \ep)
\right. \nn \\ && \left.
+\frac{256 x^2 (3-2\epsilon )}{(1+x)^4 (1+2 \epsilon )}I^{\rm RR}_6(x, \ep) 
-\frac{16  \left(1 + x^2+2 ((x-4) x+1) \epsilon \right)}{(1+x)^2 (1+2 \epsilon )}I^{\rm RR}_7(x, \ep)\right]\,,
\end{eqnarray}
where $\{I^{\rm RR}_1(x,\ep),\dots, I^{\rm RR}_7(x,\ep)\}$ are seven master integrals which must be calculated. Explicitly, we have
\begin{eqnarray}
I^{{\rm RR}}_1(x, \ep) &=& e^{2 \gamma_E  \epsilon } \pi ^{2 \epsilon - 2} \int [dk_1 \, dk_2]
\nn
\\  
I^{{\rm RR}}_2(x, \ep) &=& e^{2 \gamma_E  \epsilon } \pi ^{2 \epsilon - 2} \int [dk_1 \, dk_2] \,\frac{1}{k_1\cdot \bar{v}}
\nn
\\  
I^{{\rm RR}}_3(x, \ep) &=& e^{2 \gamma_E  \epsilon } \pi ^{2 \epsilon - 2} \int [dk_1 \, dk_2]\, \frac{1}{(k_1+k_2)\cdot v}
\nn
\\  
I^{{\rm RR}}_4(x, \ep) &=& e^{2 \gamma_E  \epsilon } \pi ^{2 \epsilon - 2} \int [dk_1 \, dk_2]\, \frac{1}{k_1 \cdot \bar{v}\, k_2\cdot v}
\nn
\\  
I^{{\rm RR}}_5(x, \ep) &=& e^{2 \gamma_E  \epsilon } \pi ^{2 \epsilon - 2} \int [dk_1 \, dk_2]\, \frac{1}{k_1 \cdot \bar{v}\, (k_1+k_2) \cdot v}
\nn
\\  
I^{{\rm RR}}_6(x, \ep) &=& e^{2 \gamma_E  \epsilon } \pi ^{2 \epsilon - 2} \int [dk_1 \, dk_2]\, \frac{k_1\cdot v}{k_1 \cdot \bar{v}\, (k_1+k_2) \cdot v}
\nn
\\ 
I^{{\rm RR}}_7(x, \ep) &=& e^{2 \gamma_E  \epsilon } \pi ^{2 \epsilon - 2} \int [dk_1 \, dk_2]\, \frac{1}{k_1 \cdot \bar{v}\, k_2 \cdot v\, (k_1-k_2)^2}\,.
\end{eqnarray}

We now go through the procedure that we used to calculate the real-virtual contributions in Section \ref{sec:realvirtual}. By making the change of variables
\begin{eqnarray}
\label{eq:RRvarchange}
I^{\rm RR}_1(x,\ep) &=& \frac{1}{(1-2 \epsilon )(1-4 \epsilon ) (3-4 \epsilon ) }F_1^{\rm RR}(x, \ep)
\nn
\\
I^{\rm RR}_2(x,\ep) &=& \frac{(1+x)}{(1-x) (1-2 \epsilon )(1-4 \epsilon ) \epsilon }F_2^{\rm RR}(x, \ep) 
\nn
\\
I^{\rm RR}_3(x,\ep) &=& \frac{(1+x)^2}{(1-x)^2 (1-2 \epsilon )^2 (1-4 \epsilon )}F_1^{\rm RR}(x, \ep) -\frac{x (1+x)}{(1-x)^3 (1-2 \epsilon )^2 \epsilon }F_3^{\rm RR}(x, \ep) 
\nn
\\
I^{\rm RR}_4(x,\ep) &=& \frac{(1+x)^2}{(1-x)^2 (1-4 \epsilon ) \epsilon ^2}F_4^{\rm RR}(x, \ep) 
\nn
\\
I^{\rm RR}_5(x,\ep) &=& -\frac{(1+x)^3}{2 (1-x) x (1-4 \epsilon ) \epsilon ^2}F_3^{\rm RR}(x, \ep) -\frac{(1+x)^3}{4 (1-x) x (1-4 \epsilon ) \epsilon ^2}F_5^{\rm RR}(x, \ep) 
\nn
\\
&-&\frac{\left(1+x^2\right) (1+x)^2}{16 (1-x)^2 x (1-4 \epsilon ) \epsilon ^2}F_6^{\rm RR}(x, \ep) 
\nn
\\
I^{\rm RR}_6(x,\ep) &=& \frac{\left(x^2-x+1\right) (1+x)^2}{(1-x)^2 x (1-2 \epsilon)^2 (1-4 \epsilon ) }F_1^{\rm RR}(x, \ep) +\frac{ (1+x)^3}{(1-x) x  (1-2 \epsilon ) (1-4 \epsilon ) \epsilon }F_2^{\rm RR}(x, \ep)
\nn
\\
&+&\frac{\left(1+x^2\right) (1+x)\left(\left(1+x^4\right) (1-(5-6 \epsilon ) \epsilon )-x^2 (2-4 (2-\epsilon ) \epsilon)\right)}{4 (1-x)^3 x^2 (1-2 \epsilon )^2 (1-4 \epsilon )  \epsilon ^2}F_3^{\rm RR}(x, \ep) 
\nn
\\
&+&\frac{\left(1+x^2\right) (1+x)^3 (1-3 \epsilon )}{8 (1-x) x^2 (1-2 \epsilon ) (1-4 \epsilon )  \epsilon ^2}F_5^{\rm RR}(x, \ep) 
+\frac{(1+x)^2}{32 (1-x)^2 x^2 (1-2 \epsilon ) (1-4 \epsilon )  \epsilon ^2}\times
\nn
\\
&\times& \Big(\left(1+x^4\right) (1-3 \epsilon )+2 x^2 \epsilon \Big)F_6^{\rm RR}(x, \ep) 
\nn
\\
I^{\rm RR}_7(x,\ep) &=& \frac{(1+x)}{(1-x) \epsilon ^3}F_7^{\rm RR}(x, \ep) \,,
\end{eqnarray}
where the $F_i^{\rm RR}(x, \ep)$ satisfy simple differential equations of the form
\begin{eqnarray} 
\label{eq:RRnormalform}
\frac{\dd}{\dd x} F_1^{\rm RR}(x, \ep) &=& 0
\nn
\\
\frac{\dd}{\dd x} F_2^{\rm RR}(x, \ep) &=& \ep \left(-\frac{1}{x}F_1^{\rm RR}(x, \ep)-\left(\frac{1}{x}+\frac{2}{1-x}\right)F_2^{\rm RR}(x, \ep)\right)
\nn
\\
\frac{\dd}{\dd x} F_3^{\rm RR}(x, \ep) &=&\ep \left(-\frac{2}{x}F_1^{\rm RR}(x, \ep) - 2 \left(\frac{1}{x}+\frac{2}{1-x}\right)F_3^{\rm RR}(x, \ep)\right)
\nn
\\
\frac{\dd}{\dd x} F_4^{\rm RR}(x, \ep) &=&\ep\left(-\frac{4}{x}F_2^{\rm RR}(x, \ep)-2\left(\frac{1}{x}+\frac{2}{1-x}\right)F_4^{\rm RR}(x, \ep)\right)
\nn
\\
\frac{\dd}{\dd x} F_5^{\rm RR}(x, \ep) &=&\ep\left(\frac{4 }{x}F_1^{\rm RR}(x, \ep)+ 4\left(\frac{1}{x}-\frac{2}{1+x}\right)F_2^{\rm RR}(x, \ep)
\right.\nn\\ 
&+& \left. 8\left(\frac{1}{x}+\frac{1}{1-x}-\frac{1}{1+x}\right)F_3^{\rm RR}(x, \ep) -2\left(\frac{1}{1-x}+\frac{1}{1+x}\right)F_5^{\rm RR}(x, \ep)
\right.\nn\\ 
&+& \left. \frac{1}{x}F_6^{\rm RR}(x, \ep)\right)
\nn
\\
\frac{\dd}{\dd x} F_6^{\rm RR}(x, \ep) &=&\ep\left(-\frac{16}{x}F_2^{\rm RR}(x, \ep)+\frac{8}{x}F_5^{\rm RR}(x, \ep)-2\left(\frac{1}{x}+\frac{2}{1-x}\right)F_6^{\rm RR}(x, \ep)\right)
\nn
\\
\frac{\dd}{\dd x} F_7^{\rm RR}(x, \ep) &=&\ep\left(\frac{4}{x}F_1^{\rm RR}(x, \ep)-\frac{2}{x}F_4^{\rm RR}(x, \ep)-2\left(\frac{1}{1-x}+\frac{1}{1+x}\right) F_7^{\rm RR}(x, \ep)\right),
\end{eqnarray}
we arrive at
\begin{eqnarray}
\label{eq:RRresultclean}
&&S_{Q\bar{Q}}^{{\rm RR}~(2)}(x,\ep) = C_F^2 \left[\vphantom{\frac{128 F_4^{\rm RR}(x, \ep) \left(x^2-(x+1)^2 \epsilon +1\right)^2}{\left(1-x^2\right)^2 \epsilon^3}}
-\frac{256 }{\epsilon }F_1^{\rm RR}(x, \ep)
+\frac{512 \left(1+x^2-(1+x)^2 \epsilon\right)}{\left(1-x^2\right) \epsilon^2}F_2^{\rm RR}(x, \ep) 
\right. \nn \\ && \left.
- \frac{128\left(1+x^2-(1+x)^2 \epsilon\right)^2}{\left(1-x^2\right)^2 \epsilon^3} F_4^{\rm RR}(x, \ep) \right]
\nn \\ 
&& + C_F n_f T_F \left[\frac{256(1-\epsilon )}{\epsilon  (1-2\epsilon)(3-2\epsilon)} F_1^{\rm RR}(x, \ep) 
-\frac{128 (1-\epsilon ) \left(1+x^2-2 (1+x)^2 \epsilon\right)}{\left(1-x^2\right)\epsilon ^2 (1-2\epsilon)(3-2\epsilon)} F_3^{\rm RR}(x, \ep)\right]
\nn \\
&& + C_A C_F \left[ -\frac{64 (\epsilon  (13-\epsilon  (9-4 (5-4 \epsilon ) (1-\epsilon ) \epsilon ))+6)}{\epsilon ^2 (1+ 2 \epsilon)^2 (1-2\epsilon)(3-2\epsilon)}F_1^{\rm RR}(x, \ep) 
\right. \nn \\ && \left.
+ \frac{32 }{\left(1-x^2\right) (1-2 \epsilon ) \epsilon ^3 (1+ 2 \epsilon)^2}\Big(-4\left(1+x^2\right) \epsilon +x^2-32 (1+x)^2 \epsilon ^5+32 (1-x)^2 \epsilon ^4
\right. \nn \\ && \left.
+ 8 (x (5 x+2)+5) \epsilon ^3-4 (x (5 x+4)+5) \epsilon ^2+1\Big)F_2^{\rm RR}(x, \ep)
\right. \nn \\ && \left.
-\frac{32 }{\left(1-x^2\right) (1-2 \epsilon ) (3-2\epsilon ) \epsilon ^3 (1+ 2 \epsilon)}
\Big(-9 \left(1+x^2\right) \epsilon -3 \left(1+x^2\right)+32 (1+x)^2 \epsilon ^5
\right. \nn \\ && \left.
-4 (x (23 x+30)+23) \epsilon ^4
+ 4 (x (19 x+3)+19) \epsilon ^3-(5-(44-5 x) x) \epsilon ^2\Big)F_3^{\rm RR}(x, \ep)
\right. \nn \\ && \left.
+\frac{8 }{\left(1-x^2\right)^2 \epsilon ^3 (1+ 2 \epsilon)^2}
\Big(-8 \left(1-x^2\right)^2 \epsilon ^3+5 \left(1+x^2\right)^2+16 (1+x)^4\epsilon ^4
\right. \nn \\ && \left.
-8 (x (10-x (4-x (x+10)))+1) \epsilon ^2
+ 2 (1-x)^2 (7-(2-7 x) x) \epsilon\Big)F_4^{\rm RR}(x, \ep)
\right. \nn \\ && \left.
+\frac{8 \left(1+x^2\right)}{\left(1-x^2\right) (1-2 \epsilon ) \epsilon ^3}F_5^{\rm RR}(x, \ep) 
+\frac{2 }{\left(1-x^2\right)^2 \epsilon ^3 \left(1-4\epsilon ^2\right)} \Big((1+x^4) (1+ 2 \epsilon)
\right. \nn \\ && \left.
+2 x^2 (1-2 (5-4 \epsilon ) \epsilon )\Big)
F_6^{\rm RR}(x, \ep)
-\frac{16 \left(1+x^2+2 (1-(4-x) x) \epsilon\right)}{\left(1-x^2\right) \epsilon ^3 (1+2 \epsilon)}
\vphantom{\frac{64 F_1^{\rm RR}(x, \ep) (\epsilon  (13-\epsilon  (9-4 (5-4 \epsilon ) (1-\epsilon ) \epsilon ))+6)}{\epsilon ^2 (1+ 2 \epsilon)^2 (1-2\epsilon)(3-2\epsilon)}}F_7^{\rm RR}(x, \ep)  \right].
\end{eqnarray}
Once again, unlike Eq.\ (\ref{eq:RRresultugly}), the above expression makes it clear that there are no power-law singularities in our result for the real-real contributions at $x = 0$.

The results given above and the input integral
\begin{equation}
I^{{\rm RR}}_1(x, \ep) = \frac{e^{2 \gamma_E  \epsilon } \Gamma^2 (1-\epsilon )}{4 \Gamma (4-4 \epsilon )}
\end{equation}
are the only ingredients that one needs in order to straightforwardly integrate the system of differential equations for the real-real contributions to arbitrarily high orders in the dimensional regularization parameter. This is the case
because, as mentioned above, regularity conditions at $x = 1$ allow for the determination of the integration constants for $I^{{\rm RR}}_2(x, \ep) - I^{{\rm RR}}_7(x, \ep)$. 
We refrain from explicitly writing our result for $S_{Q\bar{Q}}^{{\rm RR}~(2)}(x,\ep)$
to $\mathcal{O}\left(\ep^2\right)$ accuracy because of its length, but we have included it with our submission to arXiv.org as an ancillary file.

\section{R\lowercase{esults, cross-checks, and new relations}}
\label{sec:4}

In this section, we present our main result and discuss a number of consistency checks that we carried out. In fact, while considering the high-energy limit of our result, we discovered what is, to the best of our knowledge, a new class of
relations between bare soft functions in perturbative QCD. As explained in Section \ref{sec:2}, the renormalized two-loop soft function is known up to a $\mu$-independent constant, $c_2(x)$.
Needless to say, our explicit result for $\tilde{S}^{Q \bar{Q}}(x, L)$ up to and including terms of $\mathcal{O}\left(\als^2\right)$ has exactly the form predicted by Eq.\ (\ref{eq:rens}),
and we find
\begin{eqnarray}
\label{eq:c2result}
&&c_2(x) = C_F^2 \left[
2\frac{\left(1+x^2\right)^2}{\left(1-x^2\right)^2}\Bigg(G^4 (0;x)-8 G(1;x) G^3 (0;x)+8\left(\vphantom{\frac{\pi ^2}{6}}2 G^2 (1;x)+ G(0,1;x)
\right.\right. \nn \\ && \left.\left.
+\frac{\pi ^2}{6}\right) G^2 (0;x)-32\left(G(0,1;x)+\frac{\pi^2}{6}\right)G(1;x) G(0;x)+16 G^2 (0,1;x)+\frac{16\pi ^2}{3}  G(0,1;x)
\right. \nn \\ && \left.
+\frac{4 \pi^4}{9}\Bigg)+8\frac{\left(1+x^2\right)}{(1-x)^2}\left(G^2 (0;x)-4 G(1;x) G (0;x)+4 G(0,1;x)+\frac{2 \pi ^2}{3}\right)G(0;x)
\right. \nn \\ && \left.
+8\frac{(1+x)^2}{(1-x)^2}G^2(0;x)
\vphantom{\Bigg)}\right]
\nn \\ &&
+ C_F n_f T_F \left[\vphantom{\Bigg)}
-\frac{224}{27}+\frac{16(1-3 x+x^2)}{9(1-x^2)} \left( \Big(G(0;x)-4 G(1;x)\Big) G(0;x)+4 G(0,1;x)+\frac{2 \pi^2}{3}\right)
\right. \nn \\ && \left.
+\frac{8(1+x^2)}{3(1-x^2)} \Bigg(-\frac{1}{3} G^3 (0;x)+2 G(1;x) G^2 (0;x)-4\left(G^2(1;x)+\frac{\pi^2}{6}\right) G(0;x)+\frac{4\pi ^2}{3}  G(1;x)
\right. \nn \\ && \left.
+8 G(1;x) G(0,1;x)-4 G(0,0,1;x)-8 G(0,1,1;x)+4 \zeta (3)\Bigg)+\frac{16(1+30 x+x^2)}{27(1-x^2)} G(0;x)
\vphantom{\Bigg)}\right]
\nn \\ &&
+ C_A C_F \left[
-\frac{2(11 x^4-84 x^3+24 x^2+12 x-11)}{9\left(1-x^2\right)^2} G^3 (0;x)+\frac{4(55 x^2-294 x+55)}{27(1-x^2)} G(0;x)
\right. \nn \\ && \left.
+\frac{4(17 x^3-32 x^2+68 x-17)}{9(1-x^2) (1-x)} G^2(0;x)+\frac{8 \pi ^2(5 x^4+27 x^3+6 x^2+3 x+7)}{ 9\left(1-x^2\right)^2} G(0;x)
\right. \nn \\ && \left.
-4\zeta (3) \frac{\left(1+x^2\right) \left(9+x^2\right)}{\left(1-x^2\right)^2} G(0;x)+\frac{4 \pi ^2(11 x^2+66 x-25)}{27(1-x^2)}+\frac{\pi ^4\left(1+x^2\right) \left(77+221 x^2\right)}{90 \left(1-x^2\right)^2}
\right. \nn \\ && \left.
+\frac{16(26 x^2-33 x+26)}{9(1-x^2)} \Big(G(0;x) G(1;x)-G(0,1;x)\Big)+\frac{8 \pi ^2\left(1+x^2\right) \left(3+7 x^2\right)}{3\left(1-x^2\right)^2} \Big( G(0,1;x)
\right. \nn \\ && \left.
- G(0;x) G(1;x)\Big)+8\frac{3 x^3-7 x^2-5 x+1}{(1-x^2) (1-x)} \Big( G(-1;x) G^2 (0;x)-2 G(0,-1;x) G(0;x)
\right. \nn \\ && \left.
+2 G(0,0,-1;x)\Big)+16\frac{x^3-x^2+3 x+1}{(1-x^2) (1+x)} \Big(2 G(0,0,-1;x)-G(0;x) G(0,-1;x)\Big)
\right. \nn \\ && \left.
+\frac{8(13 x^4-72 x^2+11)}{3\left(1-x^2\right)^2} \Big(2 G(0,0,1;x) - G(0;x) G(0,1;x)\Big)-8\frac{\left(1+x^2\right) \left(3 x^2-1\right)}{\left(1-x^2\right)^2} \times
\right. \nn \\ && \left.
\times\Big(6 G(0,0,0,-1;x)+ G(0,-1;x) G^2(0;x)-4 G(0,0,-1;x) G(0;x)\Big)
\right. \nn \\ && \left.
+\frac{1+x^2}{1-x^2} \Bigg(\frac{8}{3} G(-1;x) G^3(0;x)+\left(\frac{4}{3} G(1;x)-8 G(0,-1;x)\right) G^2(0;x)+\bigg(24 G^2(-1;x)
\right. \nn \\ && \left.
-16 G(1;x) G(-1;x)-\frac{16\pi^2}{3} G(-1;x)-16 G(-1;x)+\frac{16}{3} G^2(1;x)-\frac{8}{3} G(0,1;x)\bigg) G(0;x)
\right. \nn \\ && \left.
-\frac{4 \pi^2}{3} G(-1;x)-\frac{100\pi^2}{9} G(1;x)-48 G(-1;x) G(0,-1;x)+16 G(1;x) G(0,-1;x)
\right. \nn \\ && \left.
+\frac{16\pi^2}{3}  G(0,-1;x)+16 G(0,-1;x)+16 G(-1;x) G(0,1;x)-\frac{32}{3} G(1;x) G(0,1;x)
\right. \nn \\ && \left.
+48 G(0,-1,-1;x)-16 G(0,-1,1;x)+\frac{8}{3} G(0,0,1;x)-16 G(0,1,-1;x)+\frac{32}{3} G(0,1,1;x)
\right. \nn \\ && \left.
+32 G(0,0,0,-1;x)\Bigg)+\frac{\left(1+x^2\right) \left(1+3 x^2\right)}{\left(1-x^2\right)^2} \bigg(\frac{8}{3} G(1;x) G^3(0;x)+16 G(0,0,1;x) G(0;x)
\right. \nn \\ && \left.
+\left(\frac{2 \pi ^2}{3}-8 G(0,1;x)\right) G^2(0;x)-16 G(0,0,0,1;x)\bigg)+\frac{x^2 \left(1+x^2\right)}{\left(1-x^2\right)^2} \bigg(-\frac{4}{3} G^4(0;x)
\right. \nn \\ && \left.
+96 G(0,0,1;x) G(0;x)-288 G(0,0,0,1;x)\bigg)+\frac{\left(1+x^2\right)^2}{\left(1-x^2\right)^2} \bigg(16 G(0,1;x) G^2(0;x)
\right. \nn \\ && \left.
+\Big(16 G(0,1,-1;x)-64 G(0,0,1;x)-32 G(1;x) G(0,1;x)+48 G(0,-1,-1;x)
\right. \nn \\ && \left.
-32 G(1;x) G(0,-1;x)+16 G(0,-1,1;x)+16 G(0,1,1;x)\Big) G(0;x)-24 G^2(0,-1;x)
\right. \nn \\ && \left.
+8 G^2(0,1;x)-\frac{4\pi ^2}{3}  G(0,-1;x)-\frac{4\pi ^2}{3}  G(0,1;x)+16 G(0,-1;x) G(0,1;x)
\right. \nn \\ && \left.
+64 G(1;x) G(0,0,-1;x)+64 G(1;x) G(0,0,1;x)-64 G(0,0,-1,1;x)+96 G(0,0,0,1;x)
\right. \nn \\ && \left.
-64 G(0,0,1,-1;x)+16 G (1;x) \zeta (3)-64 G(0,0,1,1;x)-32 G(0,1,0,-1;x)\bigg)
\right. \nn \\ && \left.
+ \frac{4\zeta(3)(13 x^4-12 x^2-49)}{3 \left(1-x^2\right)^2}+\frac{592}{27}
\vphantom{-\frac{2}{9}\frac{11 x^4-84 x^3+24 x^2+12 x-11}{\left(1-x^2\right)^2} G^3 (0;x)}\right].
\end{eqnarray}
Note that, as predicted by the non-Abelian exponentiation theorem~\cite{PHLTA.B133.90,NUPHA.B246.231}, the $C_F^2$ component of our result is precisely equal to one half the square of $c_1(x)$ modulo $C_F^2$:
\begin{equation}
c_2(x) \Big|_{C^2_F} = \frac{c_1^2(x)}{2 C_F^2} .
\end{equation}
It is worth mentioning that, in an effort to further check Eq.\ (\ref{eq:c2result}), we
did a completely independent numerical calculation of the real-virtual contributions in the spirit of reference \cite{1107.4384} and an independent construction of the unreduced integrand for the real-real contributions
by taking the appropriate limit of the full theory squared matrix element (see reference \cite{1105.0530}).

We also found it helpful to consider the threshold ($x = 1$) and high-energy ($x = 0$) limits of our results.  
In the threshold limit, the soft function must vanish by virtue of the fact that there is no phase space for soft gluon emission when the $Q \bar{Q}$ pair is produced at rest.
That this is indeed the case constitutes a non-trivial cross-check on our result because several of the real-virtual master integrals have Coulomb-like singularities at $x = 1$. In the high-energy limit,
the behavior of the soft function is significantly more complicated. This is primarily because, in the $x \to 0$ limit, logarithmic singularities in $x$ develop which correspond to 
light-like collinear singularities cut off by the mass of the heavy quark. Fortunately, a slightly more general study of this limit was carried out recently in reference \cite{1205.3662}. It is clear from the arguments given there
that the $x \to 0$ limit of our result can be obtained from a factorization formula which involves the well-known $e^+ e^- \to q \bar{q}$ soft function \cite{hep-ph/9808389}, the heavy quark fragmentation function \cite{hep-ph/0404143},
and a universal factor relating the hard function for $e^+ e^- \to q \bar{q}$ to the hard function for $e^+ e^- \to Q \bar{Q}$ in the high-energy limit~\cite{hep-ph/0612149, 0704.3582}.
With assistance, we were able to successfully compare the high-energy limit of our result to small-$x$ predictions made privately by the authors of reference \cite{1205.3662}.\footnote{We thank Ben Pecjak for his kind help in this regard.}

While studying the high-energy limit of our result, we found a completely different way to understand the $x \to 0$ limit. We observed that we could, in a rather counter-intuitive way, replace a subset of the $\ln(x)$ terms regulating the 
would-be collinear divergences in the bare $e^+ e^- \to Q \bar{Q}$ soft function by factors proportional to $1/\ep$ and thereby recover the expression for the bare $e^+ e^- \to q \bar{q}$ soft function. Quantitatively, we
found that, for all color structures not fixed by the non-Abelian exponentiation theorem, we could obtain the bare $e^+ e^- \to q \bar{q}$ soft function at $L$
loop order\footnote{The expansion coefficients of the bare $e^+ e^- \to q \bar{q}$ soft function at $L$ loop order are defined analogously to the expansion coefficients of
the bare $e^+ e^- \to Q \bar{Q}$ soft function, $S_{Q\bar{Q}}^{(L)}(x,\ep)$, introduced in Eq.\ (\ref{eq:LaplaceExp}).}, $S_{q\bar{q}}^{(L)}(\ep)$,
by making the replacements $\ln^n(x) \to 0$ for all $n > 1$ and $\ln(x) \to \frac{1}{L\ep}$ in $S_{Q\bar{Q}}^{(L)}(x\to 0,\ep)$ {\it at each order in the} $\ep$ {\it expansion};
the $\mathcal{O}\left(\ep^i\right)$ term in the $\ep$ expansion of $S_{q\bar{q}}^{(L)}(\ep)$ is thus obtained by expanding $S_{Q\bar{Q}}^{(L)}(x\to 0,\ep)$ to $\mathcal{O}\left(\ep^{i+1}\right)$
before making the above-prescribed replacements. 

To be concrete, let us illustrate the proposed correspondence for the one-loop $C_F$ and the
two-loop $C_F n_F T_F$ and $C_A C_F$ color structures. In the ancillary files containing our results for the real-virtual and real-real
contributions to $S_{Q\bar{Q}}^{(2)}(x,\epsilon)$, all expressions are given in terms of harmonic polylogarithms of argument $x$ which have been appropriately shuffled to make manifest 
the logarithmic singularities which develop in the small-$x$ limit.
It is therefore completely straightforward\footnote{The reader should be able to expand Eq.\ (\ref{eq:oneloopsol}) in $\ep$ without difficulty using the program {\tt HypExp} \cite{hep-ph/0507094}.} to write
\begin{eqnarray}
&&S_{Q\bar{Q}}^{(1)}(x\to 0,\ep) \Big|_{C_F} = -8 - 8\ln(x) +  \left[\frac{8\pi^2}{3} + 8\ln(x) + 4 \ln^2(x)\right] \ep
\\
&&+  \left[ - \frac{2\pi^2}{3}  + 16\zeta(3) - \frac{2\pi^2}{3} \ln(x) - 4\ln^2(x) - \frac{4}{3}\ln^3(x)\right] \ep^2 + \mathcal{O}\left(\ep^3\right)
\nn
\\
&& S_{Q\bar{Q}}^{(2)}(x\to 0,\ep) \Big|_{C_F n_f T_F} =  \left[\frac{32}{3} + \frac{32}{3} \ln(x)\right]\frac{1}{\ep} + \frac{160}{9} - \frac{64\pi^2}{9}  - \frac{32}{9} \ln(x) - \frac{32}{3} \ln^2(x) 
\\
&&
+ \left[ \frac{896}{27} - \frac{272\pi^2}{27} - \frac{256\zeta(3)}{3} + \left( -\frac{64}{27} + \frac{16\pi^2}{9}\right) \ln(x) + \frac{32}{9} \ln^2(x) + \frac{64}{9} \ln^3(x) \right] \ep
\nn
\\
&&
+ \left[ \frac{5248}{81} - \frac{1552\pi^2}{81} - 192\zeta(3) - \frac{32\pi^4}{135} + \left( - \frac{128}{81} - \frac{16\pi^2}{27} - \frac{448\zeta(3)}{9}\right) \ln(x)
\right.
\nn
\\
&&
\left. 
+ \left( \frac{64}{27} - \frac{16\pi^2}{9} \right) \ln^2(x) - \frac{64}{27} \ln^3(x) - \frac{32}{9} \ln^4(x)\right] \ep^2 + \mathcal{O}\left(\ep^3\right)
\nn
\\
&& S_{Q\bar{Q}}^{(2)}(x\to 0,\ep) \Big|_{C_A C_F} = \left[-\frac{88}{3} - \frac{88}{3} \ln(x)\right] \frac{1}{\ep} - \frac{392}{9} + \frac{200\pi^2}{9} - 16\zeta(3) + \left( -\frac{8}{9}  + \frac{8\pi^2}{3} \right) \ln(x) 
\nn
\\
&&
+ \frac{88}{3}\ln^2(x) + \left[ - \frac{2368}{27} + \frac{796\pi^2}{27} + \frac{1136\zeta(3)}{3} - \frac{154\pi^4}{45} + \left( -\frac{880}{27} - \frac{92\pi^2}{9} + 144\zeta(3) \right) \ln(x)
\right.
\nn
\\
&&
\left. + \left( \frac{8}{9} - \frac{8\pi^2}{3} \right) \ln^2(x) - \frac{176}{9} \ln^3(x) \right] \ep + \left[ - \frac{14240}{81} + \frac{7604\pi^2}{81} +  \frac{1936\zeta(3)}{3} 
 \right.
\\
&&
\left. + \frac{892\pi^4}{135} - 80\pi^2\zeta(3) - 640 \zeta(5)  + \left( - \frac{5216}{81} + \frac{284\pi^2}{27} - \frac{1360\zeta(3)}{9} + \frac{268\pi^4}{45} \right)\ln(x) \right.
\nn
\\
&&
\left. 
+ \left( \frac{880}{27} + \frac{92\pi^2}{9} - 144\zeta(3) \right) \ln^2(x) + \left( - \frac{16}{27} + \frac{16\pi^2}{9} \right) \ln^3(x) + \frac{88}{9} \ln^4(x)\right] \ep^2 + \mathcal{O}\left(\ep^3\right).
\nn
\label{eq:smallxQQ}
\end{eqnarray}
Using the prescription given above, it follows that
\begin{eqnarray}
\label{eq:DY1}
S_{q\bar{q}}^{(1)}(\ep)\Big|_{C_F} &=& - \frac{8}{\ep} + 2\pi^2\ep + \mathcal{O}\left(\ep^2\right)
\\
\label{eq:DY2}
S_{q\bar{q}}^{(2)}(\ep)\Big|_{C_F n_f T_F} &=& \frac{16}{3\ep^2} + \frac{80}{9\ep} + \frac{448}{27} - \frac{56\pi^2}{9} + \left[ \frac{2624}{81} - \frac{280\pi^2}{27} - \frac{992\zeta(3)}{9} \right] \ep + \mathcal{O}\left(\ep^2\right)
\\
\label{eq:DY3}
S_{q\bar{q}}^{(2)}(\ep)\Big|_{C_A C_F} &=& - \frac{44}{3\ep^2} + \left[ - \frac{268}{9} + \frac{4\pi^2}{3}\right] \frac{1}{\ep} - \frac{1616}{27} + \frac{154\pi^2}{9} + 56\zeta(3) 
\\
\nn
&+&  \left[ - \frac{9712}{81} + \frac{938\pi^2}{27} + \frac{2728\zeta(3)}{9} - \frac{4\pi^4}{9}\right] \ep + \mathcal{O}\left(\ep^2\right).
\end{eqnarray}
Indeed, Eqs.\ (\ref{eq:DY1}), (\ref{eq:DY2}), and (\ref{eq:DY3}) are completely consistent with the well-known results for $S_{q\bar{q}}^{(1)}(\ep)$ and $S_{q\bar{q}}^{(2)}(\ep)$ \cite{hep-ph/9808389}. 
To the best of our knowledge, relations between soft functions with  massive and massless Wilson lines such as the one described above have not been
discussed elsewhere in the QCD literature. Although it may not be the most efficient way to proceed in practice,
our results suggest that, to all orders in perturbation theory, one can compute a soft function with two massless soft Wilson lines directly from an analogous soft function with two
massive soft Wilson lines. It would be very interesting to investigate this further, both at higher loops and for processes with more than two soft Wilson lines.

Although we do not yet understand the origin of the novel relations that we have (presumably) discovered,
we have observed that a similar relation (with, however, $\ln(1/x)$ instead of $\ln(x)$ replaced by $1/\epsilon$ at one loop)
holds between the high-energy limit of the bare one-loop massive hemisphere soft function,\footnote{Although the massive hemisphere soft function was defined in 
reference \cite{hep-ph/0703207}, our explicit one-loop result for the bare function has not, to the best of our knowledge, appeared elsewhere in print.} 
\begin{eqnarray}
\label{eq:oneloopQQbarhemi}
S_{hemi}^{Q \bar{Q}\,~ (1)}\left(k_L, k_R, x, \mu\right) &=& \frac{4 e^{\gamma_E  \ep} \mu^{2\ep}}{\ep \Gamma (1-\ep)} \Big(k_L^{-1-2\ep}\delta\left(k_R\right)+k_R^{-1-2\ep}\delta\left(k_L\right)\Big) \times
\\ &\times&
\left(\frac{2 \ep x \Big(\,_2F_1(1,\ep-1;\ep;-x)-\,_2F_1\left(1,\ep-1;\ep;-1/x\right)\Big)}{(1-\ep)\left(1-x^2\right)}
\right. \nn \\ &+& \left.
(1-\ep) \Big(\,_2F_1\left(1,\ep;\ep+1;-1/x\right)+\,_2F_1(1,\ep;\ep+1;-x)\Big)+\ep
\vphantom{\frac{2 \ep x \Big(\,_2F_1(1,\ep-1;\ep;-x)-\,_2F_1\left(1,\ep-1;\ep;-1/x\right)\Big)}{(1-\ep)\left(1-x^2\right)}}\right),
\nn
\end{eqnarray}
and the bare one-loop massless hemisphere soft function~\cite{hep-ph/0703207}. Furthermore, 
for the correspondence between $S_{Q\bar{Q}}^{(2)}(x \to 0,\ep)$ and $S_{q\bar{q}}^{(2)}(\ep)$, we have checked that it holds for the real-virtual and real-real contributions separately,
thus implying that it cannot be explained by the known factorization properties of the $e^+ e^- \to Q \bar{Q}$  soft function in the high-energy limit alone. Altogether, the evidence strongly suggests that what we have observed is
not accidental but, rather, hints at structure in the soft limit of QCD which remains to be properly understood.

\section{O\lowercase{utlook and conclusions}}
\label{sec:5}

In this paper, we calculated the two-loop threshold soft function in perturbative QCD for heavy quark pair production at $e^+ e^-$ colliders. 
Our results will form an indispensable part of any Monte Carlo program based on the phase space slicing method which aims to provide fully differential NNLO QCD predictions for
$e^+ e^- \to Q \bar{Q}$ observables above the kinematic threshold. We refer the reader to a companion paper, reference~\cite{1408.5150}, for an explicit example of such a Monte Carlo program.
In addition, the master integrals that we computed in this paper will form a subset of the master integrals required for a complete two-loop calculation of the soft function for threshold top quark pair production at hadron colliders.
It would certainly be worthwhile and interesting to undertake this calculation and we hope to return to it in the near future. 

While performing cross-checks on our main result, we discovered interesting relations between the epsilon expansions of bare soft functions built out of
time-like Wilson lines and the epsilon expansions of bare soft functions built out of light-like Wilson lines.  
More specifically, we found that the $(n+1)$-th term in the epsilon expansion of the bare two-loop soft function considered in this work could be used, after first taking the light-like limit of the results, to straightforwardly write down the non-trivial
part\footnote{By non-trivial, we mean not constrained by the non-Abelian exponentiation theorem~\cite{PHLTA.B133.90,NUPHA.B246.231}.} of the $n$-th term in the $\ep$ expansion of the bare two-loop Drell-Yan soft function. Due to the fact that the
bare Drell-Yan soft function has poles in $\ep$ which the bare heavy quark soft function does not\footnote{The collinear divergences which are present in the Drell-Yan case are cut off by the heavy quark mass.},
it is rather surprising that such a simple correspondence exists between them. Na\"{i}vely, one might try to explain away the observed correspondence by relating it
to the known factorization properties of the heavy quark soft function discussed in reference~\cite{1205.3662}. 
However, such an approach is bound to fail because our relations can actually be used to relate the real-virtual and the real-real contributions to the two-loop soft functions in question {\it separately}.
In fact, we even observed a similar but quantitatively different phenomenon for a more exclusive type of soft function where 
the phase space available to the soft radiation is partitioned into two hemispheres by dropping a plane perpendicular to the thrust axis at the collision point.
In our opinion, the available evidence suggests that we have uncovered a new property of soft functions in perturbative QCD which demands an explanation. To this end, it would be interesting to see what happens
at higher orders in perturbation theory and to investigate whether back-to-back primary kinematics is a necessary prerequisite for there to be simple relations between pairs of soft functions in the first place.

\acknowledgments
\noindent
We are grateful to Jun Gao for useful correspondence and collaboration at an early stage of this work.
We are especially grateful to Ben Pecjak for sharing his results with us and for helping us to compare the result of our calculation in the high-energy limit to the prediction of reference \cite{1205.3662}.
HXZ and RMS gratefully acknowledge the Munich Institute for Astro- and Particle Physics for their kind hospitality during the final stage of this work. 
This research was supported in part by the Munich Institute for Astro- and Particle Physics (MIAPP) of the DFG cluster of excellence ``Origin and Structure of the Universe''.
The research of HXZ is supported in part by the US Department of Energy under contract DE–AC02–76SF00515.
The research of AvM is supported in part by the Research Center {\em Elementary Forces and Mathematical Foundations (EMG)} of the Johannes Gutenberg University of Mainz and by the German Research Foundation (DFG).
The research of RMS is supported in part by the ERC Advanced Grant EFT4LHC of the European Research Council, the Cluster of Excellence Precision Physics, Fundamental Interactions and Structure of Matter (PRISMA-EXC 1098).

\end{document}